\documentclass[useAMS,usenatbib]{mn2e}

\usepackage{amssymb}
\usepackage{amsmath}
\usepackage{times}
\usepackage{graphicx}
\usepackage{pgf}

\def\sun {\hbox{M$_{\sun}$}} 
 
\def\psr {PSR~B1822$-$09}

\def\eq{\begin{equation}}
\def\en{\end{equation}}
\def\lab{\label}
\title[X-ray/radio campaign on PSR B1822$-$09]{Simultaneous X-ray \& radio observations of the radio-mode-switching pulsar PSR~B1822$-$09}
\author[W. Hermsen et al.]{W. Hermsen$^{1,2}$, L. Kuiper$^{1}$, J.W.T. Hessels$^{2,3}$, D. Mitra$^{4,5,8}$, J.M. Rankin$^{2,5}$,
\newauthor   B.W. Stappers$^{6}$, G.A.E. Wright$^{6,7}$, R. Basu$^{4,8}$, A. Szary$^{3,8}$, J. van Leeuwen$^{2,3}$\\
$^{1}$SRON Netherlands Institute for Space Research, Sorbonnelaan 2, 3584 CA  Utrecht, The Netherlands.\\
$^{2}$Anton $P$annekoek Institute for Astronomy, University of Amsterdam, Science Park 904, 1098 XH, Amsterdam, The Netherlands.\\
$^{3}$ASTRON, the Netherlands Institute for Radio Astronomy, Postbus 2, 7990 AA, Dwingeloo, The Netherlands.\\
$^{4}$National Centre for Radio Astrophysics, (NCFRA-TIFR), Post Bag 3, Ganeshkhind, Pune University Campus, Pune 411007, India.\\
$^{5}$Physics Department, University of Vermont, Burlington, VT 05405, USA.\\
$^{6}$Jodrell Bank Centre for Astrophysics, School of Physics and Astronomy, University of Manchester, Manchester M13 9PL, UK.\\
$^{7}$Astronomy Centre, University of Sussex, Falmer, Brighton BN1 9QJ, UK.\\
$^{8}$Janusz Gil Institute of Astronomy, University of Zielona G\'ora, Lubuska 2, PL-65-265 Zielona G\'ora, Poland.}
\begin{document}

\date{Accepted 2016, November 19; Received 2016 November 17; in original form 2016 September 13}

\pagerange{\pageref{firstpage}--\pageref{lastpage}} \pubyear{2016}

\maketitle

\label{firstpage}

\begin{abstract}
We report on simultaneous X-ray and radio observations of the radio-mode-switching pulsar \psr\ with ESA's {\it XMM-Newton}
and the WSRT, GMRT and Lovell radio telescopes. 
\psr\  switches between a radio-bright and radio-quiet mode, and we discovered a relationship between the durations of its modes 
and a known underlying radio-modulation timescale within the modes. We discovered X-ray (energies 0.2$-$1.4 keV)
pulsations with a broad sinusoidal pulse, slightly lagging the radio main pulse in phase by $0.094 \pm 0.017$, with an 
energy-dependent pulsed fraction varying from $\sim$~0.15 at 0.3 keV to $\sim$~0.6 at 1 keV. No evidence is found for simultaneous 
X-ray and radio mode switching. The total X-ray spectrum consists of a cool component (T $\sim 0.96 \times 10^6$ K, hot-spot radius R $\sim 2.0$ km)
 and a hot component (T $\sim 2.2 \times 10^6$ K, R $\sim 100$ m). The hot component 
can be ascribed to the pulsed emission and the cool component to the unpulsed emission. 
The high-energy characteristics of \psr\ resemble those of middle-aged pulsars such as PSR B0656+14, 
PSR B1055$-$52 and Geminga, including an indication for pulsed high-energy gamma-ray emission in { \it Fermi} LAT data.
Explanations for the high pulsed fraction seem to require different temperatures at the two poles of this orthogonal rotator, or 
magnetic anisotropic beaming effects in its strong magnetic field.  In the X-ray skymap we found a harder
source at only $(5.1\pm 0.5)\arcsec$ from \psr, which might be a pulsar wind nebula.

\end{abstract}


\section{Introduction}

In this paper we present the results of simultaneous X-ray and radio observations of the radio-mode-switching pulsar \psr . 

The project was stimulated by the success of an earlier similar campaign involving the mode-switching pulsar 
PSR B0943+10, which resulted in the discovery of synchronous mode switching in the pulsar's radio and 
X-ray emission properties  \citep{hermsen2013}. This was then followed up by a more extended campaign \citep{mereghetti2016}
% involving XMM-Newton and the LOFAR, LWA and Arecibo radio telescopes,
in which the synchronous mode switching of PSR B0943+10 was confirmed and which enabled its nature to 
be better established, discovering that both pulsed and unpulsed X-ray emission are present in both modes 
but at differing levels and that these properties may evolve during one of the modes.

These completely unexpected results are of great importance for the understanding of the physical processes 
in pulsar magnetospheres. In addition, separate investigations have challenged the earlier perception that radio 
mode-switching is a detail of polar cap and/or magnetospheric physics by the identification of a relationship 
between the spin properties of neutron stars and their radio emission modes \citep{kramer2006, lyne2010, 
camilo2012, lorimer2012}. The implication of these results is that mode switching is due to an inherent, 
perhaps universal, pulsar process which causes a sudden change in the rate of angular momentum loss that is 
communicated along the open field lines of the magnetosphere. This link with the rapid switching observed in 
radio emission modes suggests a transformation of the global magnetospheric state in less than a rotation period. 
Nevertheless, the two X-ray radio campaigns on PSR B0943+10 have yet to conclusively answer the fundamental 
question as to whether the discovered X-ray mode switching supports a local or global magnetospheric interpretation.

To build on the PSR B0943+10 discoveries we therefore searched radio and X-ray catalogues for an attractive alternative radio 
pulsar known to exhibit mode switching with mode occurrence fractions between $\sim$ 30\% and 70\% (to 
enable statistically valid X-ray photon counts in both modes) and also reported to be an X-ray source. We did not find a 
suitable candidate there, but, searching the X-ray archives, we found that \psr\ had been in the {\it XMM-Newton}  field-of-view in a short 
observation, in which we detected the X-ray counterpart. This pulsar has a younger characteristic age of 233 kyr and a
shorter spin period $P$= 0.769 s (compared with 5 Myr and 1.1 s for PSR B0943+10). It has many properties in 
common with PSR B0943+10, but also peculiarly different characteristics.  Like PSR B0943+10, \psr\ also 
switches between a bright `B' mode and a quiet `Q' mode in its radio main pulse (MP), but on shorter times scales 
averaging a few minutes (compared with several hours for PSR B0943+10). Most remarkably, its Q-mode exhibits 
an interpulse (IP), located at about half a rotation period from the MP, which switches mode in anti-correlation with 
the MP \citep{fmw81, fw82, gil1994}. The B mode, in contrast to PSR B0943+10, exhibits a brighter but 
more complex MP as well as a precursor component (PC) some $15\degr$ longitude prior to the MP, but a barely detectable IP. 

Both PSR B0943+10 and \psr\ exhibit mode-dependent subpulse modulations, but of very different characters. 
PSR B0943+10's B-mode displays a very regular pattern of drifting subpulses with repetition period $P_3$$\approx 2P$, 
while in its Q-mode the emission is chaotic \citep{suleymanova1984, rankin2006}. In the Q-mode of \psr\ a strong 
modulation with long period   $P_{3,Q}$= (46.55$ \pm 0.88)P$ has been found, but not showing organised drifting. In its B-mode 
a weak modulation is reported with a longer and apparently harmonically-related periodicity of $P_{3,B}$= (70$\pm 3)P $
\citep{latham2012}. Further recent and detailed descriptions of its modal and pulse-sequence behaviour are given 
by \citet{backus2010, latham2012} and \citet{suleymanova2012}.

The natural interpretation of the $180\degr$ separation seen in the Q-mode between the MP and IP is 
that \psr\ is an orthogonal rotator \citep[see e.g.][]{hankins1986, gil1994, backus2010} 
with the MP and IP being produced above the magnetic poles, and both poles detected by our line of sight. 
This is an important difference to PSR B0943+10, which is believed to be rotating close to alignment ($\sim9\degr$) 
where our sight line continuously views one polar region. However, \citet{dyks2005} and \citet{malov2011, malov2013} 
pointed out that the MP-IP separation can also be explained in almost aligned rotator 
models. Given the importance of the geometry of \psr\ for the interpretation of the results of our 
X-ray/radio campaign and the absence of an exhaustive study of it in the literature, we have added such a study on 
the geometry of  \psr\ in the Appendix. There we concluded that the existing evidence strongly points to \psr\ as having 
an orthogonal rotator geometry. 
 
We carried out an X-ray/radio campaign on \psr\ with {\it XMM-Newton} and simultaneous radio observations primarily with the  
Westerbork Synthesis Radio Telescope (WSRT) at 1380 MHz, and supported by the Giant Metrewave Radio Telescope (GMRT) 
at 325 MHz and the Jodrell Bank Observatory Lovell telescope at 1420 MHz. We aimed at having two telescopes 
simultaneously  observing the pulsar to both ensure that we can identify all mode switches, and also to simultaneously 
monitor the radio characteristics at different radio frequencies.

In section 2 we present the radio observations and how we defined the radio-mode windows, and also report a 
surprising relationship
found between structures in the distributions of mode durations in the B and Q modes and their underlying 
modulation periodicities $P_3$.
In the subsequent sections we present the {\it XMM-Newton} X-ray observations (section 3), the X-ray 
Maximum-Likelihood spatial analysis of the sky maps, revealing a hard-spectrum source at a distance of only
$5\farcs1 $ from \psr\  (section 4), 
and the first detection of the X-ray pulsed signal from \psr\ folding in 
the timing analysis with an ephemeris derived from Lovell telescope radio monitoring of \psr\ (section 5). 
In section 6 we present the search for X-ray mode switching using a combined spatial and 
timing analysis.
The spectral characterisation of the total, pulsed and steady unpulsed X-ray emissions is performed in section 7, 
followed by a summary of our findings in section 8 and a discussion of them in section 9. Finally, our overall conclusions are presented in section 10.

\section{Radio observations and results}

\begin{table}
\caption{\bf Jodrell Bank ephemeris of \psr\ valid during our {\it XMM-Newton} observations.}
\label{ephemeris}
\begin{center}
\begin{tabular}{c c}
\hline
\hline
Right Ascension (J2000)   & $ 18^{\hbox{\scriptsize h}} 25^{\hbox{\scriptsize m}} 30\fs630(5) $ \\
\vspace{-2.5mm}\\
Declination (J2000)    &   $ -9\degr 35\arcmin 22\farcs12(3)$   \\
\vspace{-2.5mm}\\
Epoch (TDB)    & 55836  \\
\vspace{-2.5mm}\\
$\nu ~(Hz)$& 1.30036814629(1)   \\
\vspace{-2.5mm}\\
$\dot{\nu}~(Hz~s^{-1}) $  & $-8.87283(2) \times  10^{-14}$\\
\vspace{-2.5mm}\\
$\ddot{\nu}~(Hz~s^{-2}) $  & $-5.6(1) \times 10^{-25}$  \\
\vspace{-2.5mm}\\
 Start (MJD) & 54923  \\
\vspace{-2.5mm}\\
 End (MJD) &  56749\\
\vspace{-2.5mm}\\
 Solar System Ephemeris  &  DE200 \\
\hline \\
\end{tabular}
\end{center}
Numbers in parenthesis are the 1-$\sigma$ errors on the least significant digits.
\end{table}

\subsection{WSRT radio observations and analysis}

We observed \psr\ with the Westerbork Synthesis Radio Telescope (WSRT)
using the tied-array mode, in which the individual signals of each dish are
coherently summed to a single synthesised array beam by applying
appropriate time and phase delays.  Twelve of the
fourteen\footnote{The rest were unavailable because of maintenance and
  the transition of the telescope to a new receiver system.} WSRT 25-m
dishes participated in the observations, and the PuMaII backend
\citep{kss08} was used to record baseband voltage data from $8 \times
20$-MHz subbands together spanning frequencies of $1300-1460$\,MHz.
We converted the baseband data to both folded (10-s sub-integrations)
and single-pulse archives using {\tt
  dspsr}\footnote{http://dspsr.sourceforge.net/manuals/dspsr/}
\citep{sb11} and an up-to-date ephemeris (see Table~\ref{ephemeris}) provided by ongoing timing of
\psr\ with the Lovell telescope \citep{hobbs2004}.  Using tools from the 
{\tt PSRCHIVE}\footnotemark[2]
software suite \citep{hsm04,sdo12}, the data were cleaned of radio
frequency interference (RFI) and the 8 20-MHz bands were combined into
a single data cube of pulse phase, radio frequency and intensity
(Stokes I).  WSRT provided data during all (eight in total) {\it XMM-Newton} observing
sessions, and covered practically all of the relevant X-ray windows
(see Tables 2 and 3).

\begin{table}
\caption{{\bf Radio observations of \psr\ in 2013 and 2014
with the WSRT, GMRT and Lovell telescope, together covering the simultaneous observations with 
{\it XMM-Newton} (Table~\ref{table_xmm_obs}).}}
\label{radio_obs}
\begin{center}
\begin{tabular}{c c c c c c}
\hline
\hline
Telescope & Date & Freq. &Start (UT) & End (UT) & duration \\
  &   yyyy-mm-dd  & MHz  & hh:mm        &    hh:mm & hours    \\
\hline
\hline

WSRT &    &    &   &    \\
      & 2013-09-10 & 1380 & 14:52 &   22:22 & 7.49   \\
\vspace{-2.5mm}\\
       & 2013-09-18 & 1380 & 13:01  &  15:04 & 2.06   \\
 \vspace{-2.5mm}\\
       & 2013-09-18 & 1380 & 15:15  &  20:31 & 5.25   \\      
\vspace{-2.5mm}\\
   & 2013-09-22 & 1380 &  12:56 &  20:45 & 7.82 \\
\vspace{-2.5mm}\\
    & 2013-09-28 &1380 &  12:49 &  20:18 & 7.50   \\
\vspace{-2.5mm}\\
   & 2013-09-30 & 1380 &  13:17 &  21:36 & 8.33  \\
\vspace{-2.5mm}\\
    & 2013-10-06 &  1380 & 12:22 & 19:12 & 6.83   \\
\vspace{-2.5mm}\\
    & 2014-03-10 & 1380  &  06:22 &  10:22 & 4.00   \\
  \vspace{-2.5mm}\\
    & 2014-03-10 & 1380  &  10:26 &  11:56 & 1.50   \\  
\vspace{-2.5mm}\\
   & 2014-03-12 &  1380 &  06:34& 10:34 & 4.00   \\
\vspace{-2.5mm}\\
   & 2014-03-12 &  1380 &  10:38 & 11:48 & 1.16   \\  
\hline
GMRT &    &    &    &    \\
       & 2013-09-10 & 338 & 13:15  &  18:42  & 5.45\\
  \vspace{-2.5mm}\\
       & 2013-09-18 & 624 & 17:35  &  18:12   & 0.62\\            
\vspace{-2.5mm}\\
   & 2013-09-22 & 624 &  11:37 &  18:06 & 6.48\\
\vspace{-2.5mm}\\
    & 2013-09-28 & 624 &  09:32 &  17:30   & 7.95\\
\vspace{-2.5mm}\\
   & 2013-09-30 &  624 &  09:36 &  17:20  & 7.73\\
\vspace{-2.5mm}\\
    & 2013-10-06 & 624  & 10:41 & 17:01   & 6.33 \\
\hline   
 Lovell  &   &    &    &   &  \\
    & 2013-09-10 & 1420 & 14:55 &   23:45   &  8.83\\
\vspace{-2.5mm}\\
       & 2013-09-18 & 1420 & 17:05  &  19:47  &  2.70\\
\vspace{-2.5mm}\\
   & 2013-09-22 & 1420 &  13:31 &  20:57 & 7.43\\
\vspace{-2.5mm}\\
    & 2013-10-06 & 1420  & 14:30 & 19:13  &  4.72\\
\vspace{-2.5mm}\\
   & 2014-03-12 & 1420  &  09:56 & 12:05  &  2.15\\
\vspace{-2mm}\\
\hline \\
\end{tabular}
\end{center}
\end{table}

%%%%%%%%%%%%%%%%%%%%%%%%%%%%%%%%%%%%%%%%%%%%%%%%%%%%%%%%%%%%%%%%%%%%%%%%%%%%%%%%%%%%%%%%%%%%%%%%%%%%%%%%%%%%%%%%%%%%%%%%%%%%%%%%

\begin{figure}
  \begin{center}
     \includegraphics[width=70mm,angle=-90]{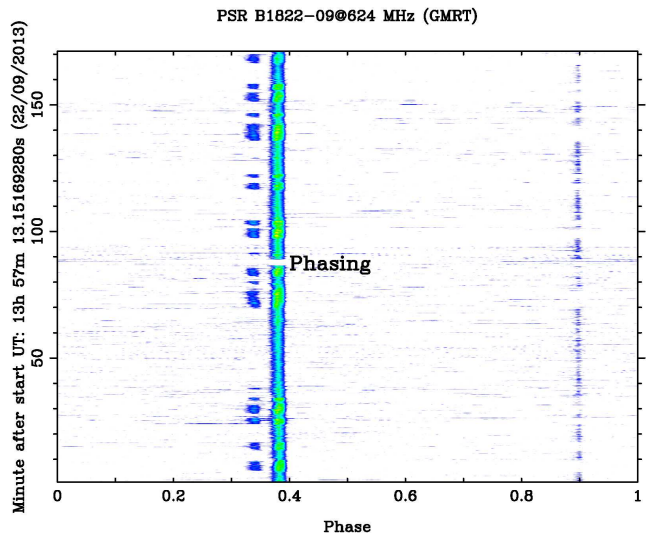}
     \caption{\label{GMRT-PSR1822-09} GMRT observation at 624 MHz of \psr , 22 September 2013, showing single 
     pulse sequences. Observation time vs. pulsar phase: at phase $\sim$ 0.34 the precursor (PC) switches on/off; at phase $\sim$ 0.38 
     the main pulse (MP) switches between the bright (B) and weaker quiet (Q) mode; at phase $\sim$ 0.9 the interpulse (IP) switches on/off 
     in anti-correlation with the MP and PC. There is some RFI as well, and that corresponds to the other features.
     Phasing: time in which there was no available data on \psr\ because the telescope needed to point to a calibration source.}
  \end{center}
\end{figure}
%%%%%%%%%%%%%%%%%%%%%%%%%%%%%%%%%%%%%%%%%%%%%%%%%%%%%%%%%%%%%%%%%%%%%%%%%%%%%%%%%%%%%%%%%%%%%%%%%%%%%%%%%%%%%%%%%%%%%%%%%%%%%%%%

\subsection{GMRT observations}

Simultaneous \psr\ observations were carried out at the Giant
Meterwave Radio Telescope (GMRT) for the first six observing sessions
in September and October 2013, at 338 MHz on the first session and 624
MHz for the rest of the sessions (see Table~\ref{radio_obs}).  The GMRT is a multi-element
aperture synthesis telescope \citep{swarup1991} consisting of 30
antennas distributed over a 25-km-diameter area which can be
configured as a single dish in both coherent and incoherent array
modes of operation.  The observations discussed here used the coherent
(or more commonly called Ôphased arrayÕ) mode \citep{gupta2000, sirothia2000}. 
At these observing frequencies, the GMRT is equipped
with dual linear feeds which are converted to left and right-handed
circular polarisations via a hybrid. The dual polarisation signals are
passed through a superheterodyne system and down converted to the
baseband which are finally fed to the GMRT software backend \citep{roy2010}. 
In the backend the FX corelator algorithm is
implemented and in the phased array mode the voltage signals from all
antennas are added in phase, which is finally recorded as a total
power time series. We used a total bandwidth of about 33 MHz spread
over 256 channels with time resolution of 0.122 ms.

The observing sessions at GMRT were such that only partial overlap,
mostly during the earlier part of the observations, between WSRT,
Lovell and {\it XMM-Newton} were possible. Even within
the overlap region some parts of the data were not usable due
to RFI or technical problems. At GMRT frequencies the strength of the
interpulse is much more prominent than seen at higher frequencies, and
hence the alternate emission between the IP and the PC for the Q and B
modes, respectively, could be easily verified in these data. In
regions of overlap with the WSRT observations, the mode switches were seen to correlate extremely
well with what was derived based on the WSRT data.  Fig.~\ref{GMRT-PSR1822-09}
shows as an example of the moding behaviour of \psr\ single pulse sequences
measured with the GMRT at 624 MHz covering part of one of our {\it XMM-Newton} observations.
The MP and PC are shown to simultaneously switch on/off on short time scales (typically minutes), while the
IP switches on/off in anti-correlation with the MP and PC.

%%%%%%%%%%%%%%%%%%%%%%%%%%%%%%%%%%%%%%%%%%%%%%%%%%%%%%%%%%%%%%%%%%%%%%%%%%%%%%%%%%%%%%%%%%%%%%%%%%%%%%%%%%%%%%%%%%%%%%%%%%%%%%%%

\begin{figure*}
  \begin{center}
     \includegraphics[width=178mm]{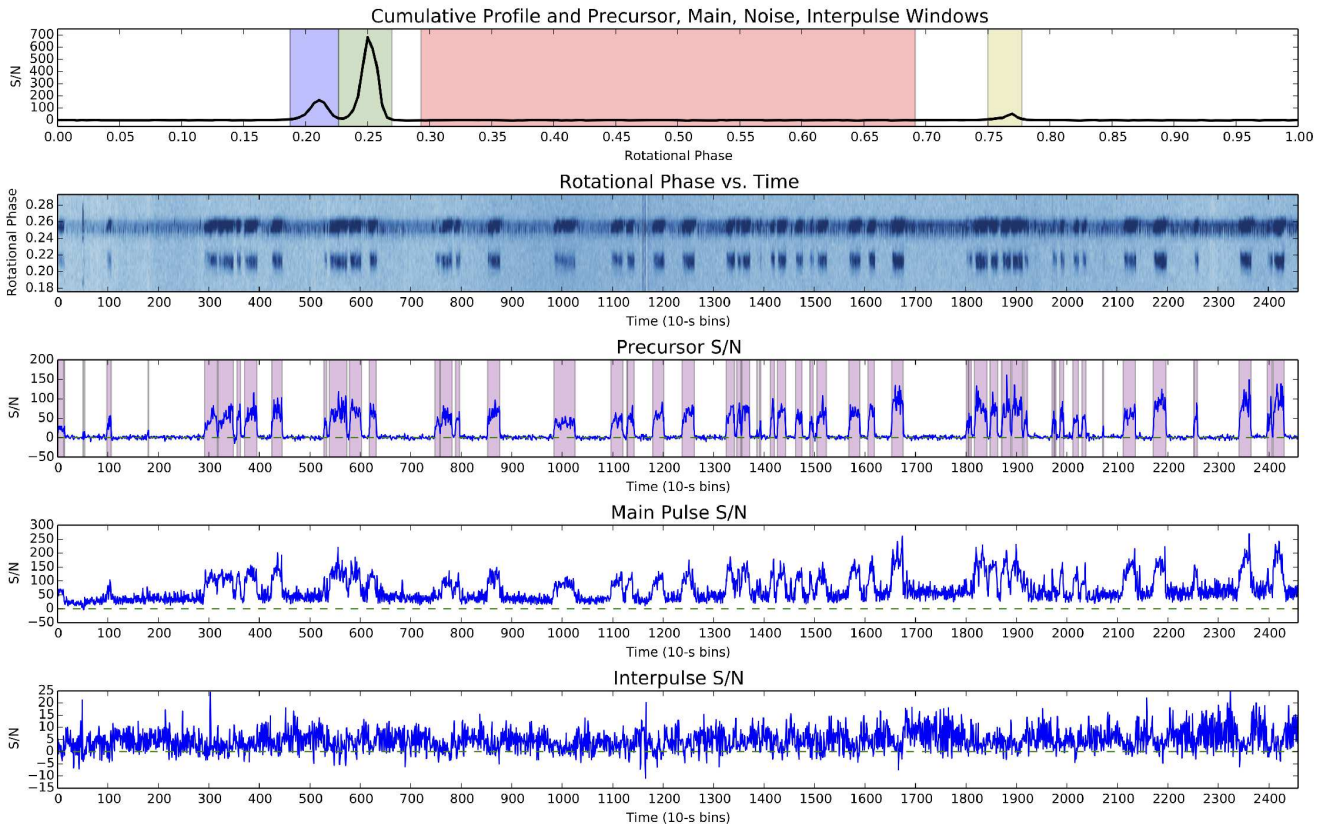}
     \caption{\label{WSRT-modes-PSR1822-09} WSRT observation at 1380 MHz of \psr\ for 6.8 hours, 6 October 2013. 
     From top to bottom: cumulative pulse profile with indicated in colour the PC (purple), MP (green), off-pulse region / ``Noise'' (orange) 
     and IP (yellow) phase windows, followed by  a zoom-in on the PC and MP as a function of time; then, the signal-to-noise ratios (S/N) of the detections in time bins of 10s of the PC, 
     the MP and the IP. The x-axis in the bottom 4 panels gives the number of 10-s bins. The PC exhibits a very clear on/off switching, and has been used to define the bright 
     (B, pink) and quiet (Q) mode intervals.}
  \end{center}
\end{figure*}
%%%%%%%%%%%%%%%%%%%%%%%%%%%%%%%%%%%%%%%%%%%%%%%%%%%%%%%%%%%%%%%%%%%%%%%%%%%%%%%%%%%%%%%%%%%%%%%%%%%%%%%%%%%%%%%%%%%%%%%%%%%%%%%%

\subsection{Lovell Telescope observations} 

We observed PSR B1822-09 using the 76 m Lovell Telescope at the Jodrell Bank Observatory on five occasions (Table~\ref{radio_obs}) simultaneous with the {\it XMM-Newton} observations 
(Tables~\ref{table_xmm_obs}). 
Observations were made using a dual-polarisation cryogenic receiver at a central frequency of 1520\,MHz with a total of about 380\,MHz of useful 
bandwidth after removal of known and intermittent radio-frequency interference. The incoming analogue data was sampled, dedispersed and folded 
in real-time using two different pulsar backends, the digital filter bank (DFB) and the coherent dedispersion backend called the ROACH \citep{bassa2016}. 
The DFB data filtered the bandwidth into 0.5\,MHz wide channels and incoherently dedispersed and folded into 10-s sub-integrations using the 
same ephemeris as used for the WSRT observations. The ROACH data were coherently dedipsersed and two data products were generated: 
folded profiles with 10-s sub-integrations with 1\,MHz wide channels and single-pulse archives with 4\,MHz wide channels. The coherent 
dedispersion used the {\tt  dspsr} \citep{sb11} software and post-processing made use of the {\tt PSRCHIVE}
software suite \citep{hsm04,sdo12}. 

The Lovell telescope data overlapped with almost the entire WSRT data set and the {\it XMM-Newton} observing windows and so was used as an 
independent check of the detailed analysis of the WSRT data described below. A series of randomly chosen mode transitions were 
identified in the Lovell data, using either, or both, of the single pulse and 10-s subintegration data and compared with those transitions 
identified in the WSRT data. These confirmed the times of the mode transitions and because of the different offsets between the start 
times of the 10-s sub-integrations and single pulses could be used to estimate the maximum number of pulses which might be 
attributed to the incorrect mode.

\subsection{Mode determinations using WSRT, Lovell \& GMRT data}

\label{modes}

In our analysis we make the basic assumption that, at any given time,
\psr\ is either emitting in the B or Q mode.  While this assumption 
seems to be strictly true for most mode-switching pulsars 
and was formerly thought to be true for \psr, \citet {latham2012} found 
evidence for short intervals of mixed modes in \psr\ in the vicinity of the 
transitions from one mode to the other with a duration of a few pulses. 
These intervals, however, are sufficiently 
short that they have little effect on our analysis.  

Mode determination was based primarily on the WSRT data set because it
provided the most complete overlap with the {\it XMM-Newton} sessions.
The Lovell and GMRT data provided checks of the WSRT results, and the
strategy of using 3 radio telescopes in parallel was to provide
redundancy in the case of technical problems at one of the
observatories during one of the {\it XMM-Newton} sessions (fortunately
no such issues affected the general observing campaign).  

Using WSRT data, we defined for each observation pulse-phase windows encompassing
\psr's pulse PC, MP, IP and a reference
off-pulse region (Figure~\ref{WSRT-modes-PSR1822-09}).  For each 10-s sub-integration
(equivalent to $\sim 13$ pulse periods of $P = 0.769$\,s) we measured
the S/N of the PC, MP and IP (Figure~\ref{WSRT-modes-PSR1822-09}).
We found that setting a S/N threshold on the precursor emission
provided a robust and automated method of separating the B and Q-mode intervals --
despite the low number of pulses per 10-s sub-integration. The
presence of PC emission is in anti-correlation with the presence of
the IP, but, the relative weakness of the IP at 1.4\,GHz
observing frequencies makes it a comparatively less reliable mode
indicator. While \psr\ can sometimes switch modes on $<$ 10-s timescales, such switches were not resolvable 
using this method. These intervals, however, are sufficiently short that they have negligible 
effect on the results of the X-ray analysis.

The first radio observations took place in Sept/Oct 2013 and comprised six sessions of between 7 and 
8.5 hours long (see Table~\ref{radio_obs}), virtually contiguous with the times of the {\it XMM-Newton} observations listed in 
Table~\ref{table_xmm_obs}, amounting to a total of 162.620 ks (211,500 pulses). 
On the successive dates the Q-mode percentage was  65.1\%, 62.4\%, 60.6\%, 62.9\%, 66.6\%, 66.9\%, 
with an overall figure of 64.1\%.  The second set of observations occurred on March 10 and 12 in 2014 (Table~\ref{table_xmm_obs}). 
There were two sequences on each day, and the short second sequence of both had to be truncated 
(after 4 ks and 2.7 ks respectively) as the pulsar was setting and the mode resolution poor. 
Altogether the useful observations amounted to 35.340 ks (46,000 pulses). The total Q-mode percentage 
of these observations was  63.5\%, giving a composite figure of 63.9\% for 2013 and 2014 together.

The total number of mode changes was 952, implying an average of one mode change every three and 
a half minutes (208 s).  The average Q-mode sequence was 270 s ($\approx$ 347 pulses), B-mode sequence  150 s
($\approx$195 pulses). The longest Q-mode was 4420 pulses (3400 s), recorded on 28 Sept, and the longest B-mode 
was 971 pulses (670 s) in the same observation and directly after the aforementioned long mode. 
Note that in a single 8-hr observation (i.e. comparable 
with each of our observations) \citet{latham2012} found examples of both Q- and 
B-mode lengths greater than those recorded here. An explanation for this possible discrepancy might be that some
of our long mode windows are interrupted by a short (single 10-s integration) switch to the other mode that could in some
cases be spurious and caused by RFI. For example, in Figure~\ref{WSRT-modes-PSR1822-09} there is a very short ``B-mode'' instance that breaks
up what would be an otherwise longish Q-mode sequence. Alternatively, such short mode interruptions might have been overlooked
by \citet{latham2012}.

It is relevant to note that \citet{lyne2010} studied the spin down of \psr\ in a sample of 17 pulsars over $\sim$ 20 years.
The value of the $\dot{\nu}$ of \psr\ averaged over typically 50 days was found to change between two well-defined values, correlated with
changes of the average pulse profile: for the high-$\mid\dot{\nu}\mid$ state, the precursor 
is weak and the interpulse is strong, with the reverse occurring for the smaller-$\mid\dot{\nu}\mid$ state. This correlation of 
large changes in $\dot{\nu}$ with pulse shape changes, found for a substantial number of pulsars, suggests a change in magnetospheric
particle current flow. We verified that during all eight observing sessions of our campaign, the average $\dot{\nu}$ values of \psr\ were similar, 
not in  a high-$\mid\dot{\nu}\mid$ state.

\subsection{Mode-length Histograms}

\begin{figure}
\begin{center}
\includegraphics[width=9.0cm,angle=-0]{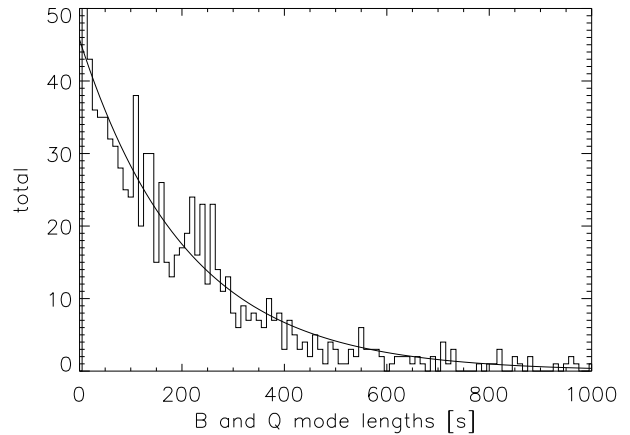}  

\caption{The histogram of the length of the intervals between the mode changes of PSR B1822-09. 
The data combines all observations in 2013 and 2014. The plotted exponential curve  is the 
expectation if the mode changes occurred at random with a probability 0.048 per 10-s interval.}

\label{modewindows_total}
\end{center}
\end{figure}

Fig.~\ref{modewindows_total}  shows a histogram of the lengths of the intervals between the approximately 952
mode changes observed in 55 hours, sampled every 10s. If these were occurring at random it would imply a 
probability of 952/(55$\times$60$\times$6)=0.048 per 10~s. The associated exponential 
distribution describing the outcome of a Poisson process would be
\eq
\lab{1}
N(x)=952\times0.048\exp(-0.048x)
\en
Note that this curve closely matches the observations apart from the spike in the first single bin of 10 s (truncated in the figure), which contained 98 modes.

%%%%%%%%%%%%%%%%%%%%%%%%%%%%%%%%%%%%%%%%%%%%%%%%%%%%%%%%%%%%%%%%
%%%%%%%%%%%%%%%%%%%%%%%%%%%%%%%%%%%%%%%%%%%%%%%%%%%%%%%%%%%%%%%%%

%%%%%%%%%%%%%%%%%%%%%%%%%%%%%%%%%%%%%%%%%%%%%%%%%%%%%%%%%%%%%%%%%%

\begin{figure}
\begin{center}
{\includegraphics[width=9.0cm,angle=0]{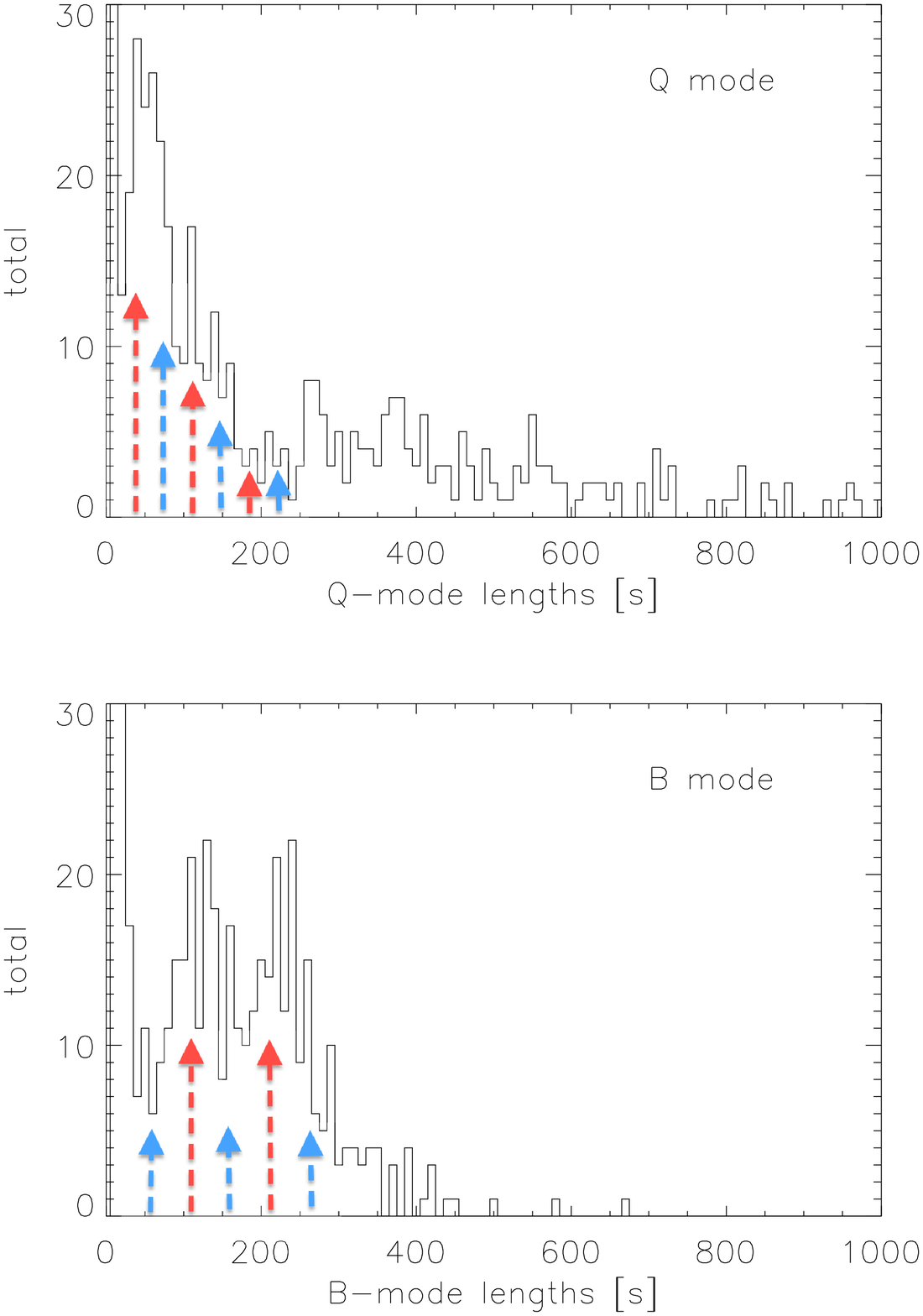}}
\caption{Upper figure: the histogram of the  Q-mode lengths. Lower figure: the equivalent histogram for B-mode.
The red and blue lines show successive multiples of the $P_3$ for each mode reported by \citet{latham2012}. Note that red 
lines (odd multiples of $P_3$ for Q, even for B) tend to coincide with maxima, and blue lines (even multiples 
of $P_3$ for Q, odd for B) with minima. Since the $P_3$s of Q and B (35.8 s and 53.8 s respectively) 
are resonant (2:3) the red lines at $3\times{35.8}= 2\times{53.8}=107.4$ s and the red/blue lines at $2\times107.4=215.4$ 
s are common to both histograms.}
\label{Q_B_modewindows}
\end{center}
\end{figure}

%%%%%%%%%%%%%%%%%%%%%%%%%%%%%%%%%%%%%%%%%%%%%%%%%%%%%%%%%%%%
%%%%%%%%%%%%%%%%%%%%%%%%%%%%%%%%%%%%%%%%%%%%%%%%%%%%%%%%%%%%

Fig.~\ref{Q_B_modewindows}({\it upper}) shows the histogram of the $\approx{952/2=476}$ Q-mode lengths. In addition to 
the ultra-short modes of less than 20 s, we note a narrow peak at short lengths and a long trailing tail, suggesting the 
possibility of two kinds of Q-mode sequences (or three if we include the ultra-short). Fig.~\ref{Q_B_modewindows}({\it lower}) shows the histogram 
of the B-mode lengths. Apart from the spike in the first 10 s, it is noticeable that the distribution has two equal peaks 
(at $\sim$ 120 s  and $\sim$ 240 s) with a shallow dip between them. There is also a much shorter tail towards longer mode 
durations than in the Q-mode distribution. 

Both the histograms of Fig.~\ref{Q_B_modewindows} have been overlaid with vertical red and blue lines which mark 
multiples of the central modulation periodicities ($P_3$) found by \citet{latham2012} in 325 MHz observations of long Q- and 
B-mode sequences. Thus $P_{3,Q}$, expressed in terms of the rotation period $P$, is 46.55$P$ (35.8 s) for the 
Q-mode and $P_{3,B}=70P$ (53.6 s) for the B-mode, close to a ratio of 2:3. The markers appear to coincide with features in the histograms.

In Fig.~\ref{Q_B_modewindows}({\it upper}) for the Q-mode the first (red) line closely aligns with the initial peak of the distribution, 
implying that the most common length for a Q-mode sequence, if one ignores the larger peak at the shortest time scales, 
is the same as the modulation timescale found in 
longer Q-mode sequences. It also suggests that all Q-mode sequences start at about the same phase of modulation.

%%%%%%%%%%%%%%%%%%%%%%%%%%%%%%%%%%%%%%%%%%%%%%%%%%%%%%%%%%%%%%%%%

\begin{figure}
\begin{center}
\includegraphics[width=9.0cm,angle=-0]{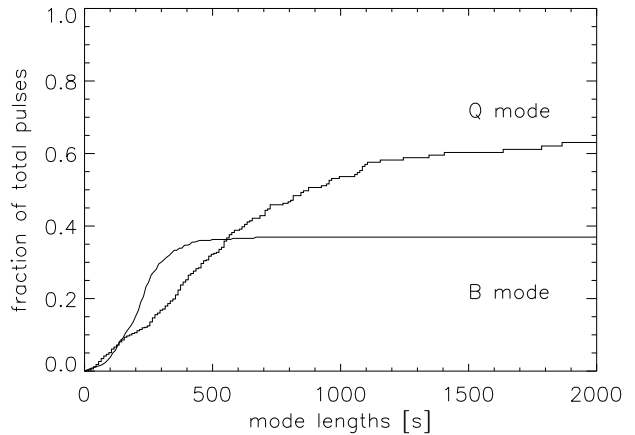}  

\caption{Cumulative distributions of the fraction of the total number of pulses, in separately the Q and B mode, as a function of mode length of PSR B1822-09.}

\label{cumulative_lengths}
\end{center}
\end{figure}

%%%%%%%%%%%%%%%%%%%%%%%%%%%%%%%%%%%%%%%%%%%%%%%%%%%%%%%%%%%%%%%%%%
We previously noted that both Figures 3\&4 exhibit initial peaks for mode lengths $\le$ 20~s and remarked earlier that mode switches on $<$ 10-s timescales were not resolvable.
In addition, RFI can mimic such short mode lengths and could be responsible for the initial peaks. In order to investigate how significant these short mode-lengths might be for our analysis, we generated
cumulative distributions of the fraction of the total number of pulses ($\sim$ fraction of total observation time)
in the Q and B mode as a function of mode length (see Fig.~\ref{cumulative_lengths}). This figure shows that the contribution of these short-lengths intervals amounts to
$<$ 1\% of the total number of pulses, thus can be ignored
in the following discussion of the radio characteristics, as well as for the accumulation of X-ray events in the Q and B modes.

In the Q mode, at multiples of $P_{3,Q}$ little can be discerned as peaks or minima in the histogram, possibly because 
the decline in the height of the histogram is rapid and the bin size will smooth any features. However, a clear 
minimum appears around $6\times{P_{3,Q}}=215.2$ s. For this reason we have given the $P_{3,Q}$ lines alternating 
colours of red and blue, speculating that there may be an unresolvable underlying periodicity of $2\times{P_{3,Q}}$, 
corresponding to what is found in the B-mode. 

In the B-mode histogram of Fig.~\ref{Q_B_modewindows}({\it lower}) a \emph{minimum} appears at the $P_{3,B}=70P=53.8$ 
s (blue line), so B-mode sequences of this length are rare in the observations. 
Overwhelmingly, they are found near the first two multiples of $2\times{P_{3,B}}$, i.e. 107.6 s and 215.2 s 
(red lines), separated by a further minimum at $\sim$ $3\times{P_{3,B}}=161.4$ s (blue line). This strongly 
supports the suggestion by \citet{latham2012} that the precursor PC is modulated at $2\times{P_{3,B}}$ and is weak where the MP is 
weak \citep[see their Figure 12]{latham2012}. As with the Q-mode, the presence of peaks and troughs in the histogram implies 
that the B-mode starts at roughly the same phase of its modulation.

Note that the cycle of maxima and minima in the two modes complement one another, with maxima 
in one roughly coinciding with minima in the other. In fact the B-mode maximum and Q-mode minimum 
at $\sim$ 215.2 s ($=280P$) precisely coincide since 1/280 is half the beat frequency of 1/46.55 and 1/70.

The important conclusion to be drawn from Fig.~\ref{Q_B_modewindows} is that for the first time in a mode-switching pulsar 
a clear yet subtle relationship has been found between the duration of its modes and a known underlying 
modulation timescale of the intensities 
within the modes. It is all the more remarkable that in Fig.~\ref{modewindows_total} the combined histogram of both modes suggests -- 
possibly coincidentally -- 
an overall random distribution of the mode lengths. 

\section{{\it XMM-Newton} X-ray observations}

Before our work, only upper limits were published on X-ray emission from \psr\ 
\citep{alpar1995, slane1995}. However, we found in the archives of {\it XMM-Newton} a short observation with
an effective on-source exposure of 4.8 ks. We clearly detected the source at a significance level of 6.9$\sigma$ (energies 0.2 -- 2 keV) 
with a count rate slightly higher than what we had measured for the time-averaged count rate from PSR B0943+10, enabling us to propose this
campaign. Later, \citet{prinz} also analysed this observation and found the source as well.

We obtained six {\it XMM-Newton} X-ray observations of \psr\ in September and October 2013, and two in March 2014 with durations 
between $\sim$ 6 and 9 hours (see Table~\ref{table_xmm_obs}). In this work we used only data obtained with the EPIC instrument 
aboard {\it XMM-Newton}, which consists of one camera based on Pn CCDs \citep{struder2001}, the Pn, and two cameras based on 
MOS CCDs \citep{turner2001}, MOS~1 and MOS~2. During all observations, the Pn camera was operated in Large Window mode, 
which provides a 
time resolution of 47.7 ms, and the MOS cameras were operated in Small Window mode with a time resolution of 0.3 s. For the 
three cameras we used the thin optical filter. The EPIC Pn count rate in the 10-12 keV energy range was always significantly below 1.2 
counts s$^{-1}$,
indicating that our observations were not affected by increased particle backgrounds due to soft proton flares.
The total lifetime (dead time corrected exposure) for the Pn, MOS~1,
and MOS~2  cameras are given in Table~\ref{lifetimes}.

\begin{table}
\caption{{\bf {\it XMM-Newton} observations of \psr\ in 2013 and 2014}}
\label{table_xmm_obs}
\begin{center}
\begin{tabular}{c c c c}
\hline
\hline
Obs. ID    & Date & Start time Pn/MOS& End time Pn/MOS \\
            &    yyyy-mm-dd  & hh:mm (UT)        &    hh:mm (UT)     \\
\hline
\hline
0720730901      & 2013-09-10 & 15:41 / 15:12 &   22:07 / 22:07    \\
\vspace{-2.5mm}\\
0720731001        & 2013-09-18 & 13:51 / 13:21  &  19:43 / 19:43   \\
\vspace{-2.5mm}\\
0720731101       & 2013-09-22 & 13:41 / 13:11 &  20:35 / 20:34 \\
\vspace{-2.5mm}\\
0720731201       & 2013-09-28 & 13:39 / 13:10 &  19:31 / 19:31   \\
\vspace{-2.5mm}\\
0720731301      & 2013-09-30 & 13:48 / 13:19 &  21:34 / 21:34  \\
\vspace{-2.5mm}\\
0720731601      & 2013-10-06 & 13:18 / 12:32 & 19:10 / 19:09   \\
\vspace{-2.5mm}\\
0720731401      & 2014-03-10 & 07:37 / 07:08 &  13:29 / 13:29   \\
\vspace{-2.5mm}\\
0720731501    & 2014-03-12 & 07:30 / 07:01& 16:59 / 16:59   \\

\vspace{-2mm}\\
\hline \\
\end{tabular}
\end{center}
\end{table}

\begin{table}
\caption{{\bf Total lifetime (dead time corrected exposure) of our {\it XMM-Newton} observations of 
\psr\ for the MOS~1, MOS~2 and Pn cameras, and the corresponding lifetimes in the radio 
Q and B-mode time intervals. The radio observations covered about 87\% of the total {\it XMM-Newton} observations.}}
\label{lifetimes}
\begin{center}
\begin{tabular}{c c c c}
\hline
\hline
Camera   & Total & Q-mode & B-mode \\
   &    ks   &   ks &  ks\\
\hline
\hline
MOS~1     & 200,652 & 113,494 &   62,664   \\
\vspace{-2.5mm}\\
MOS~2     & 200,387 & 113,254  &  62,620   \\
\vspace{-2.5mm}\\
Pn    & 178,101 & 98,888 &  56,312 \\
\vspace{-2mm}\\
\hline \\
\end{tabular}
\end{center}
\end{table}

\section{X-ray spatial analysis}
\label{spatial}

In the spatial analysis we apply a two-dimensional Maximum Likelihood (ML) method using our knowledge of the two-dimensional 
point-source
signature (the point spread function; PSF), and taking into account the Poissonian nature of the counting process. The PSF is fitted 
to the measured two-dimensional count distribution on top of a background structure, assumed to be flat, at the known position of the 
source or on a grid of positions in the source region when the source position is not known. Typical fit-region sizes are of the order 
of 30-60\arcsec. 

At first the events\footnote{Each event $i$ is characterised 
by its (barycentred) arrival time $t_i$, spatial coordinates 
$x_i, y_i$ , energy $E_i$, event pattern $\xi_i$ and flag $\digamma_i$. We used $\xi_i$ = [0,4] and flag $\digamma_i$ = 0 for both Pn and MOS.} are sorted in counts skymaps with two-dimensional
pixels typically of size 2\arcsec $\times$ 2\arcsec. To obtain the number of source counts the following 
quantity, 
$L=\ln\left( \prod_{i,j}\ (\mu_{ij}^{N_{ij}} \exp(-\mu_{ij}) /N_{ij}!)\right) = \sum_{i,j} N_{ij}\ln(\mu_{ij}) - \mu_{ij} -\ln(N_{ij}!) $, 
where $\mu_{ij}=\beta+\sigma \cdot {PSF}_{ij}$ is the expectation value for pixel $(i,j)$ and $N_{ij}$ the number of counts measured 
in pixel $(i,j)$, is maximised  simultaneously with respect to the background parameter $\beta$ and source scale parameter $\sigma$. 
The PSF is normalised to unity. Therefore, the total number of source counts  $\sigma$ (background free) is automatically obtained.
The second derivative matrix of $L$ evaluated at the optimum contains information about the uncertainties on the derived parameters, 
$\beta$ and $\sigma$.

%%%%%%%%%%%%%%%%%%%%%%%%%%%%%%%%%%%%%%%%%%%%%%%%%%%%%%%%%%%%%%%%%%%%%%%%%%%%%%%%%%%%%%%%%%%%%%%%%%%%%%%%%%%%%%%%%%%%%%%%%%%%%%%%

\begin{figure*}
  \begin{center}
     \includegraphics[width=8cm,height=8cm,angle=0,bb=15 15 185 185,clip=]{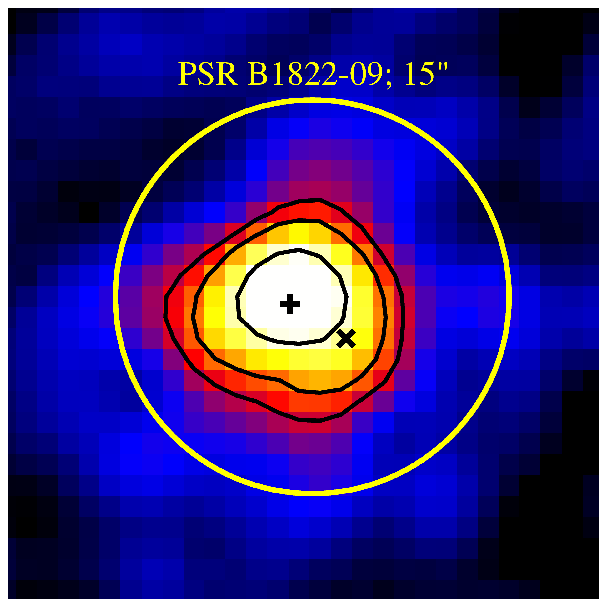}
     \includegraphics[width=8cm,height=8cm,angle=0,bb=15 15 185 185,clip=]{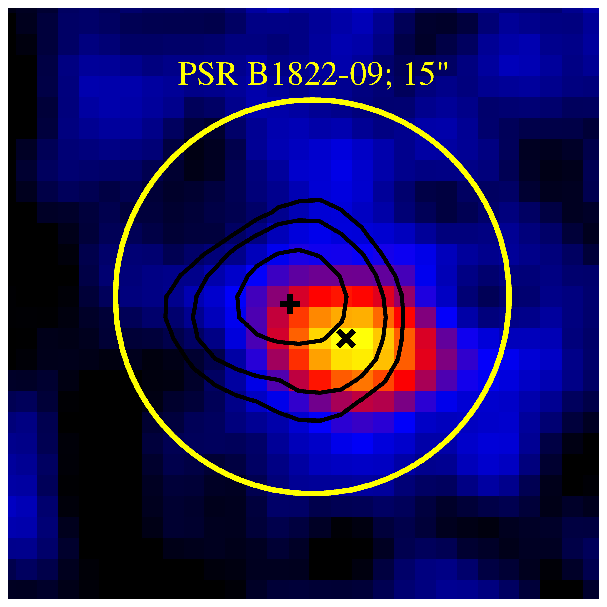}
     \caption{\label{MapsMOS1-2} {\it XMM-Newton} MOS~1\&2 Images for energies 0.2-1.4 keV (left) and 1.4-10 keV (right), combining 
     data from all eight observations. The yellow circles are centered on the radio position of \psr\ with radii $15\arcsec $.
     The position uncertainty from radio timing is  $\ll 1 \arcsec$, see Table 1. The black 
     contours represent contours of the image for energies 0.2-1.4 keV, shown in both images for comparison with the colour images. 
     The X-ray position of \psr\ (+; with mean astrometric error of  $(1.25 \pm 0.65) \arcsec$ and negligible statistical uncertainty) 
     and that of a nearby hard source at an angular separation of $(5.1\pm 0.5)\arcsec$ (x) are indicated.  Below 1.4 keV the central 
     excess is consistent with the presence of one source, positionally consistent with \psr ; above 1.4 keV also with only one source, 
     the nearby hard source.}
  \end{center}
\end{figure*}
%%%%%%%%%%%%%%%%%%%%%%%%%%%%%%%%%%%%%%%%%%%%%%%%%%%%%%%%%%%%%%%%%%%%%%%%%%%%%%%%%%%%%%%%%%%%%%%%%%%%%%%%%%%%%%%%%%%%%%%%%%%%%%%%

We applied this approach to the EPIC Pn and MOS~1\&2 data (treating these separately), and obtained a very significant  source 
excess at the position of \psr . However, it turned out that this excess is not consistent with a single point source at the position of \psr .
This is already visible in the raw count maps, as is illustrated in Fig. \ref{MapsMOS1-2}, which shows sky maps combining data from 
all eight observations with MOS~1\&2 for energies 0.2 $-$1.4 keV (left) and 1.4 $-$ 10 keV (right). 
Below 1.4 keV the central excess is consistent 
with the presence of one source, positionally consistent with that of \psr\  and with a detection significance of 32.6 sigma (with 796+/- 38 counts assigned to the pulsar). 
From our X-ray analysis we derived 
the following accurate position coordinates of \psr:
$\alpha_{2000} = 18^{\hbox{\scriptsize h}} 25^{\hbox{\scriptsize m}} 30\fs726 , \delta_{2000} = -9\degr 35\arcmin 23\farcs14$. 
The mean astrometric error for {\it XMM-Newton} is  $(1.25 \pm 0.65) \arcsec$ and the statistical uncertainties are negligible.
The X-ray position is within $1\farcs8$ from the VLBI radio position of \citet{fomalont1992} and at $\sim1^{\prime \prime}$ from the Lovell timing position in Table~\ref{ephemeris}.
Also above 1.4 keV an excess is visible,  but now shifted in position, with a detection significance of 14.4 sigma (with 268 +/- 26 counts),
and source position 
$\alpha_{2000} = 18^{\hbox{\scriptsize h}} 25^{\hbox{\scriptsize m}} 30\fs429$ , $\delta_{2000} = -9\degr 35\arcmin 25\farcs76 $. 
The sources are thus separated by $(5.1\pm 0.5)\arcsec$. 
The source positions are indicated in the figure. The independent EPIC Pn maps confirm these results.

\subsection{Nature of nearby hard source? }

The ML spatial analysis showed that the X-ray distribution can be explained with the presence of two point sources. However, given the $6\arcsec$ FWHM PSF
of {\it XMM-Newton} and the measured separation of $(5.1\pm 0.5)\arcsec$, we cannot exclude that the neighbouring  excess is slightly extended.
Of 13 nearby middle-aged pulsars with {\it XMM-Newton} and/or {\it Chandra} X-ray observations 6 have prominent X-ray Pulsar Wind Nebulae (PWNe), 2 have very faint PWNe 
and for 4 there are no detections reported yet in current data \citep{posselt2015}.
For example, the most well-known nearby middle-aged pulsar Geminga exhibits a very prominent PWN, discovered by \citet{caraveo2003} and more recently studied in most detail
by \citet{pavlov2010} exploiting deep {\it Chandra} observations. The latter authors confirmed that the Geminga PWN has three tail-like components with patchy structures,
including three possibly moving blobs. The spectra of the PWN elements appeared to be rather hard with photon index $\sim$1. Our new source might be such a hard-spectrum
element of a PWN of \psr, but a deep high-spatial resolution Chandra observation is required to resolve the ambiguity of whether we are detecting in X-rays an extended structure or a compact object.

To investigate this further, we searched for a counterpart at lower energies 
by scrutinising the optical/IR archives. We found deep (almost) on-source exposures performed on
June 15 and July 20, 2004 by the ESO 3.6 m NTT equipped with the $5\farcm 5 \times 5\farcm 5$ CCD 
imaging camera SUSI2. We downloaded several images taken through different filters and correlated 
these with the UCAC3 catalogue for astrometric calibration yielding accurate positions at $0\farcs 1$ level. 
In the 60-s, 600-s and 800-s exposures (R-, H$_{\alpha}$ and V-bands, respectively) we found no candidate
counterpart at the location of the ÔhardÕ source, nor at the position of PSR B1822-09, down to a limiting magnitude in the V-band of  $\sim 25.0$, 
adopting an input spectrum of a A0 V star.

Independent of its nature, the discovery of this nearby hard source requires in our further analysis, notably when source 
spectra are derived, that both sources are fitted simultaneously for energies above 1.4 keV. For the {\it XMM-Newton} PSF and the low
counting statistics of the new source, it suffices to assume a point-source structure.

%\begin{table}
%\caption{{\bf Detection significances ($Z_1^2$ value) of the X-ray pulse of \psr\ in differential energy intervals}}
%\label{pulse_significances}
%\begin{center}
%\begin{tabular}{c c}
%\hline
%\hline
%Energy interval    & Detection significance  \\
 % keV          &    Number of $\sigma$'s     \\
%\hline
%\hline
%0.40 $-$ 0.51     & 2.0    \\
%\vspace{-2.5mm}\\
%0.51 $-$ 0.66  & 5.0   \\
%\vspace{-2.5mm}\\
%0.66 $-$ 0.85   & 4.9 \\
%\vspace{-2.5mm}\\
%0.85 $-$ 1.09     & 5.4   \\
%\vspace{-2.5mm}\\
%1.09 $-$ 1.40    & 3.7  \\
%\vspace{-2mm}\\
%\hline \\
%\end{tabular}
%\end{center}
%\end{table}

%%%%%%%%%%%%%%%%%%%%%%%%%%%%%%%%%%%%%%%%%%%%%%%%%%%%%%%%%%%%%%%%%%%%%%%%%%%%%%%%%%%%%%%%%%%%%%%%%%%%%%%%%%%%%%%%%%%%

\begin{figure*}
  \begin{center}
     \includegraphics[width=88mm]{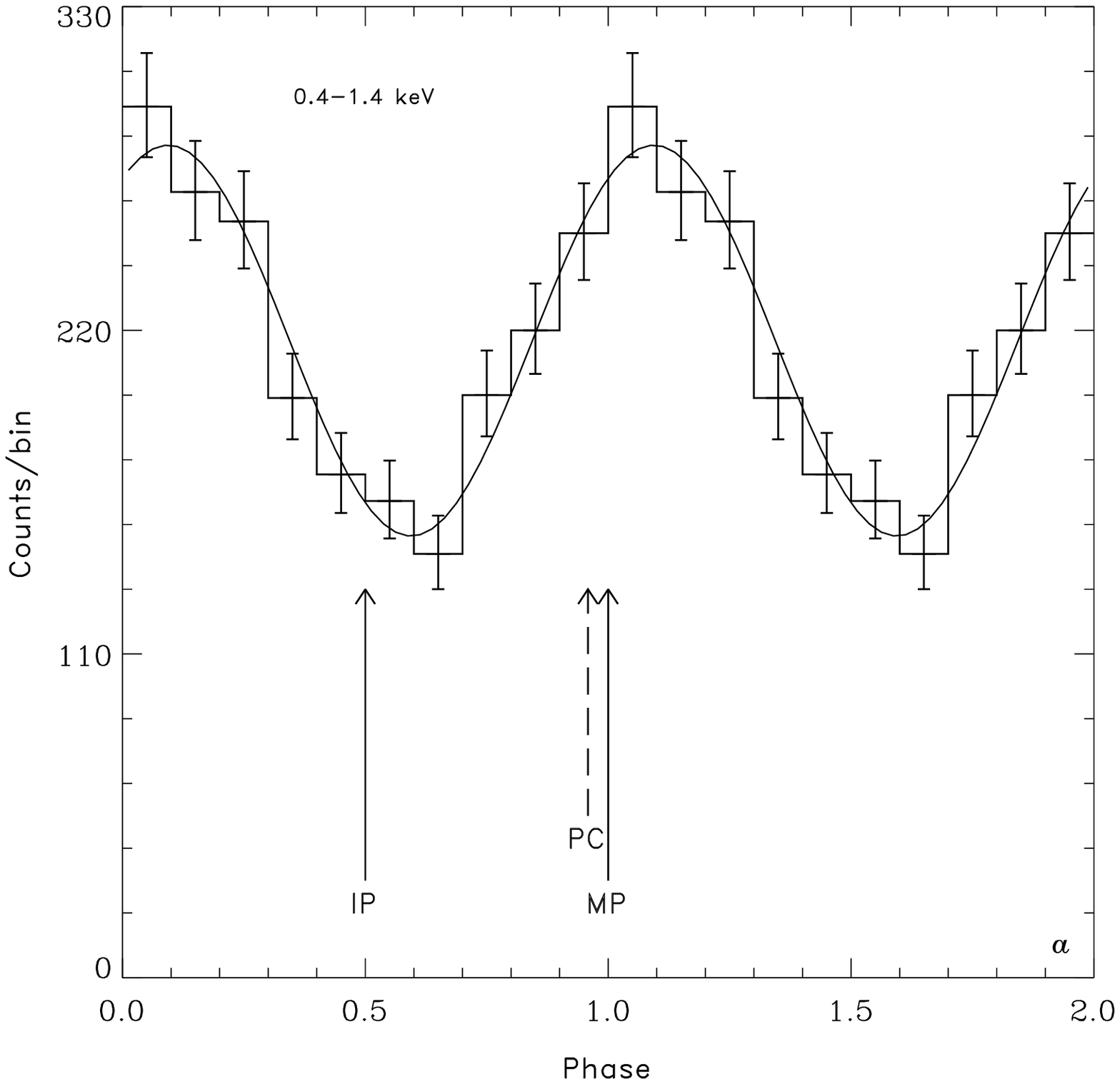} 
     \includegraphics[width=88mm]{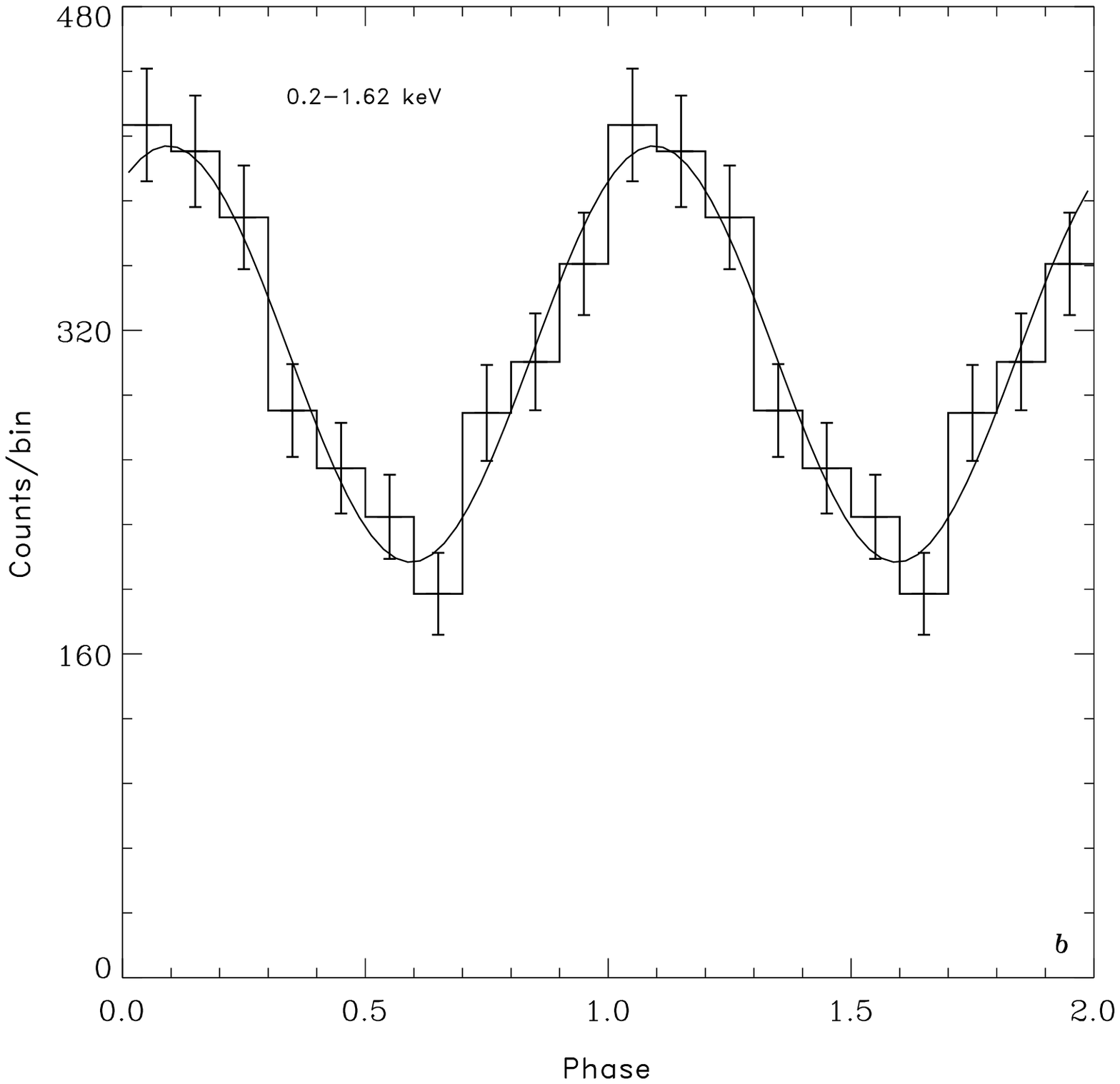} 
     \caption{\label{Profile1822} Pulse profile of \psr , using data from {\it XMM-Newton} Pn and MOS~1\&2 detectors 
     for all 8 observations. {\it Left, (\bf a)}, derived by phase folding of event arrival times (see text) for energies 0.4-1.4 keV selecting
     only events within an aperture of $20^{\arcsec}$; 
     detection significance $9.6 \sigma$. {\it Right, (\bf b)}, derived by phase-resolved imaging (see text) for 
     energies 0.2-1.62 keV, including all detected source events in the counts map; 
     sky background has been modelled out. The radio main pulse peaks at phase 0. The solid-line profiles show fits 
     with sinusoids peaking at phase 0.094 $\pm$ 0.017. The arrows in Fig~\ref{Profile1822}{\it a} mark the phases of the
     radio main pulse (MP), interpulse (IP) and precursor (PC).}
  \end{center}
\end{figure*}
%%%%%%%%%%%%%%%%%%%%%%%%%%%%%%%%%%%%%%%%%%%%%%%%%%%%%%%%%%%%%%%%%%%%%%%%%%%%%%%%%%%%%%%%%%%%%%%%%%%%%%%%%%%%%%%%%%%%%%%%%%%%%%%%

%%%%%%%%%%%%%%%%%%%%%%%%%%%%%%%%%%%%%%%%%%%%%%%%%%%%%%%%%%%%%%%%%%%%%%%%%%%%%%%%%%%%%%%%%%%%%%%%%%%%%%%%%%%%%%%%%%%%%%%%%%%%%%%%

\begin{figure*}
  \begin{center}
    \includegraphics[width=150mm]{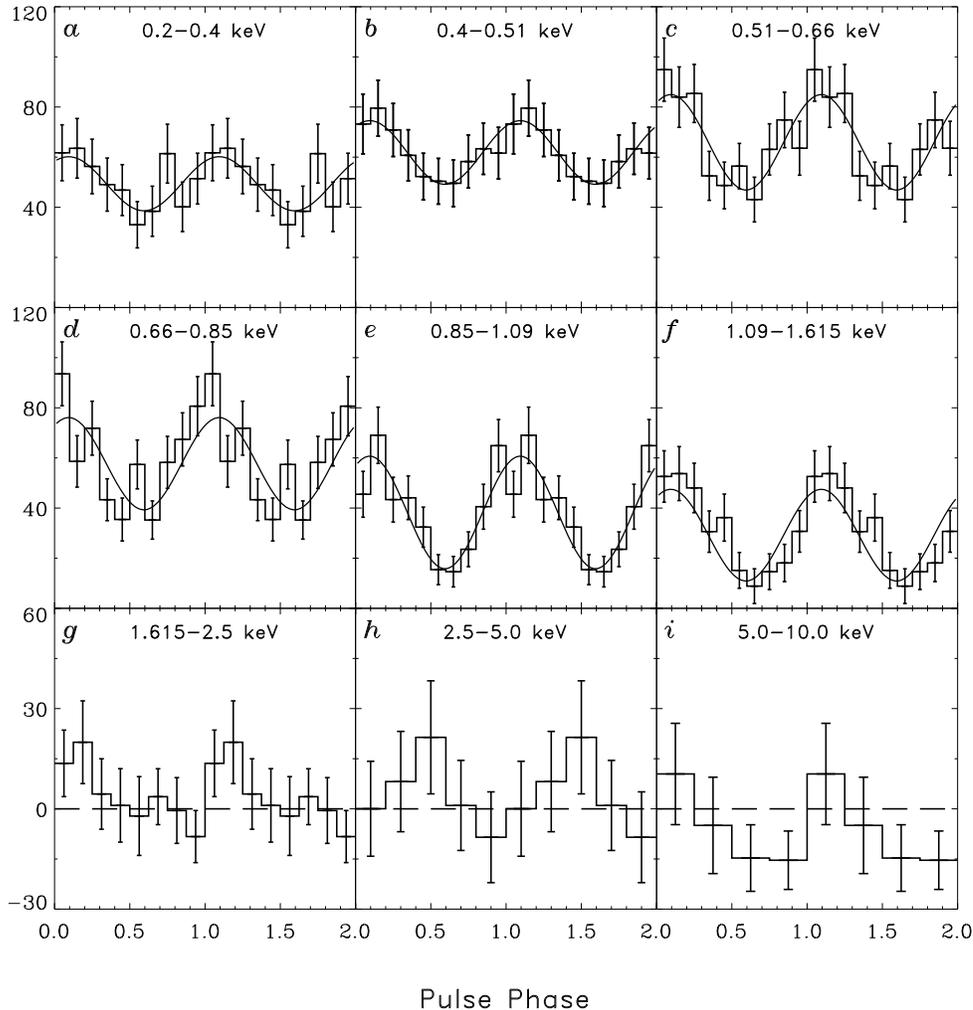}
    \caption{\label{Profiles1822spatial}  Pulse profiles of \psr\ in differential energy intervals 
     between 0.2 and 10 keV, obtained from phase-resolved spatial analyses using 
     {\it XMM-Newton} Pn and MOS~1\&2 data for all observations listed in Table~\ref{table_xmm_obs}. The solid-line profiles show fits with a 
     sinusoid centred on the phase of the X-ray pulse for the integral 0.4-1.4 keV energy interval. The radio main pulse peaks 
     at phase 0. The celestial background, assumed to be flat, is modelled out. 
     The y-axis gives per phase bin the derived number of pulsed plus unpulsed counts from the point source.} 
  \end{center}
\end{figure*}
%%%%%%%%%%%%%%%%%%%%%%%%%%%%%%%%%%%%%%%%%%%%%%%%%%%%%%%%%%%%%%%%%%%%%%%%%%%%%%%%%%%%%%%%%%%%%%%%%%%%%%%%%%%%%%%%%%%%%%%%%%%%%%%%

%%%%%%%%%%%%%%%%%%%%%%%%%%%%%%%%%%%%%%%%%%%%%%%%%%%%%%%%%%%%%%%%%%%%%%%%%%%%%%%%%%%%%%%%%%%%%%%%%%%%%%%%%%%%%%%%%%%%

\begin{figure}
  \begin{center}
     \includegraphics[width=84mm]{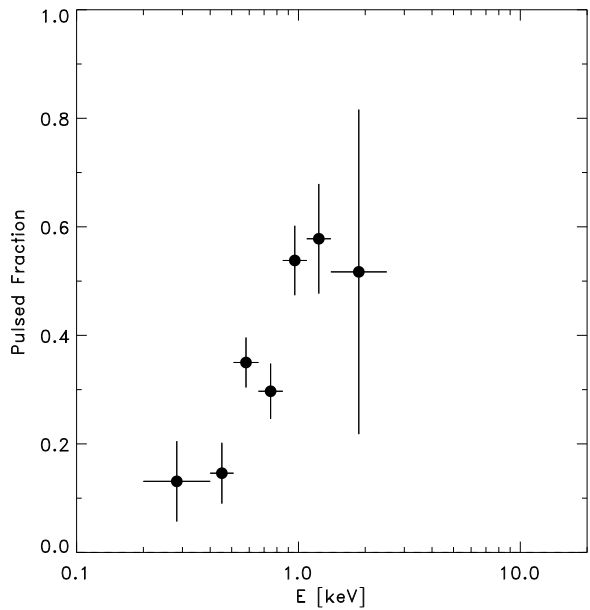} 
     \caption{\label{Pulsedfractions} \psr: pulsed fraction as a function of energy as derived
     from a three dimensional ML analysis (see text). The energy dependence shown, indicates that 
     between 0.2 and 2 keV the spectrum of the pulsed emission is harder than the spectrum of the unpulsed emission. }
  \end{center}
\end{figure}
%%%%%%%%%%%%%%%%%%%%%%%%%%%%%%%%%%%%%%%%%%%%%%%%%%%%%%%%%%%%%%%%%%%%%%%%%%%%%%%%%%%%%%%%%%%%%%%%%%%%%%%%%%%%%%%%%%%%%%%%%%%%%%%%

\section{X-ray timing analysis}
\label{timing}

We performed timing analyses by selecting Pn and MOS~1\&2 events detected within a $20^{\prime \prime}$ aperture around the X-ray position of \psr .
The times of arrival were converted to arrival times at the Solar System Barycentre and folded upon the ephemeris of \psr\  given in 
Table~\ref{ephemeris}. Fig. \ref{Profile1822}{\it a} shows the discovery of pulsed X-ray emission from \psr\ with the detection 
of a broad pulse in the energy band 0.4-1.4 keV at a significance of $9.6 \sigma$ \citep[$Z_1^2$ value]{buccheri1983}. 
The measured X-ray profile can be well
fitted with a sinusoid and reaches its maximum at phase $0.094 \pm 0.017$, with the radio main pulse, which is also sinusoidal in shape, peaking at phase 0 
(indicated in Fig. \ref{Profile1822}{\it a}). The radio precursor precedes the radio main pulse, and the X-ray pulse lags the radio. Furthermore, 
we do not see a pulse or local maximum at the phase of the interpulse, something we might expect for the geometry of an orthogonal rotator.
However, if there is a weaker X-ray pulse at the phase of the interpulse, also sinusoidal in shape, the summed profile with the main pulse  
would hide the X-ray interpulse and result in a phase distribution as shown in 
Fig. \ref{Profile1822}{\it a}. Therefore we cannot state at this stage of the analysis, whether the measured distribution is due
to two broad X-ray pulses, a strong one peaking close to the phase of the main pulse, and a weaker one peaking around the phase of the interpulse,
or that we measure an unpulsed level of steady emission with a superposed single pulse component. In the 
spectral analysis we will address these two options.

The pulsar-phase distribution in Fig. \ref{Profile1822}{\it a} contains, in addition to pulsar-source 
counts (pulsed plus unpulsed), a flat celestial background level. This background level can be suppressed by applying 
phase-resolved spatial analysis: for each phase bin count sky maps are produced and with the two-dimensional ML method 
the number of source counts is determined per phase bin. This approach gives the  distribution shown in Fig. \ref{Profile1822}{\it b}
for the integral energy range 0.2 $-$ 1.62 keV. All counts detected from \psr\ (pulsed plus unpulsed) are contained 
in this pulsar-phase distribution. Therefore, the number of source counts is much larger than in Fig. \ref{Profile1822}{\it a},
because the latter contains only source counts (plus sky-background counts) detected within a $20^{\prime \prime}$ aperture around the \psr\ position,
while the two-dimensional tail of the PSF extends well beyond the applied aperture. As a result, the energy range over which the pulse profile is detectable,
widens somewhat.

We applied the phase-resolved spatial analysis (all Pn and MOS~1\&2, observations) in nine 
differential energy intervals between 0.2 and 10 keV. Fig.~\ref{Profiles1822spatial} shows the resulting pulse-phase distributions.  
From 0.2 to 1.6 keV between $\sim$ 300 to 700 source counts (pulsed plus unpulsed) are detected per energy interval, 
while above 1.6 keV the distributions are consistent with zero counts from \psr\ (the zero level is indicated by the broken line).
This figure also shows that there is no evidence for a variation of the pulse shape with energy over 
the 0.2 to 1.6 keV band. Namely, for all phase distributions below 1.6 keV fits with sinusoids, with maximum at the same phase as
shown in Fig.~\ref{Profile1822}, above flat unpulsed source-count levels nicely represent the data, as is shown in the figure. 
Fig.~\ref{Profiles1822spatial} shows already by eye how 
the relative contributions of pulsed and unpulsed emission vary with energy. 

In Fig.~\ref{Profiles1822spatial} it is apparent that the pulse profile is sinusoidal across the {\it XMM-Newton} energy band. 
We can therefore generalize the ML method by taking into account also the pulse-phase information of 
the events (three-dimensional approach), by assuming an energy-independent shape of the pulse profile. 
Sorting the events now according to their spatial (x,y) and pulse phase $\phi$ values we can write for the 
expectation value 
of bin (i,j,k): $\mu_{ijk}=\beta+\sigma_u \cdot {PSF}_{ij} + \sigma_p \cdot {PSF}_{ij} \cdot \Phi_k$.
In this formula the value of the normalized pulse profile at bin $k$ is represented by $\Phi_k$, 
while $\sigma_u$ and $\sigma_p$ 
correspond to the unpulsed and pulsed component scale factors. From $\sigma_u$ and $\sigma_p$ 
the pulsed fraction $\eta$ can be 
determined as $\eta=1/(1+ (N_{\phi} \cdot \sigma_u/\sigma_p))$, with $N_{\phi}$ the number of bins of the normalized pulse profile.

Applying the three-dimensional ML method we derived the pulsed fraction as a function of energy, shown in Fig.~\ref{Pulsedfractions}. 
At low energies of $\sim$ 0.3 keV the pulsed fraction starts at a level of
$\sim$ 0.15 and reaches a value of $\sim$ 0.6 around 1 keV. This clearly shows that the X-ray spectrum 
of the pulsed signal is much harder than that of the unpulsed emission. In \S~\ref{pulsed_unpulsed_spectra} we will derive 
the corresponding spectral parameters.

%%%%%%%%%%%%%%%%%%%%%%%%%%%%%%%%%%%%%%%%%%%%%%%%%%%%%%%%%%%%%%%%%%%%%%%%%%%%%%%%%%%%%%%%%%%%%%%%%%%%%%%%%%%%%%%%%%%%%%%%%%%%%%%%

\begin{figure}
 \begin{center}
     \includegraphics[width=82.5mm]{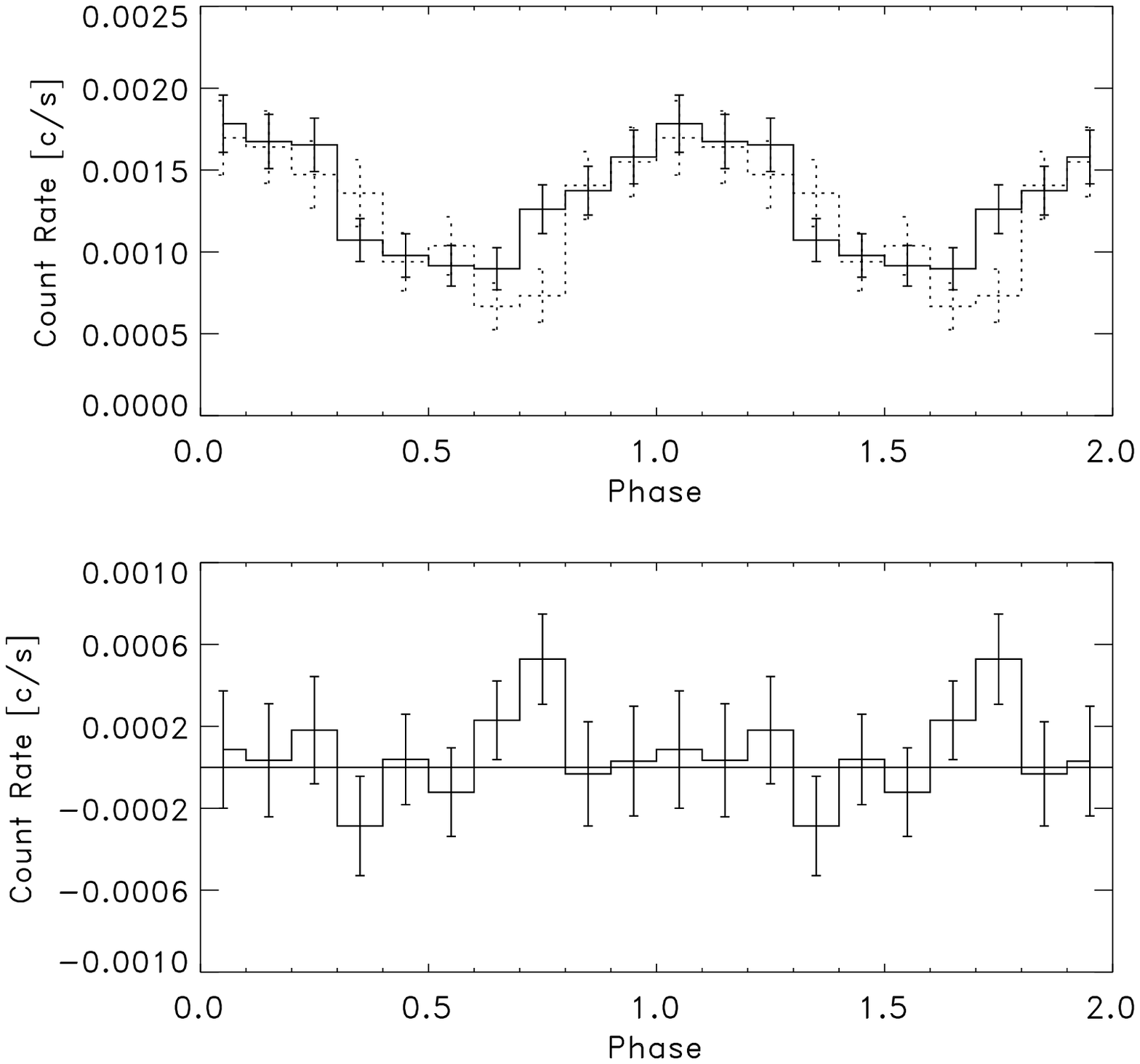}     
     \caption{\label{Modeswitch} Pulse profiles of \psr\ obtained from phase-resolved spatial analyses 
     using Pn and MOS~1\&2 data. The celestial background is modelled out. 
     Events are selected in the radio Q-mode (solid lines)  or B-mode (broken lines) time 
     windows for energies 0.4-1.4 keV.  The pulse is detected in the Q and B mode. 
     The probability that the two profiles are drawn from the same parent distribution is $97.5 \% $. 
     The difference between the two mode-selected profiles is shown in the lower panel.}
  \end{center}
\end{figure}

%%%%%%%%%%%%%%%%%%%%%%%%%%%%%%%%%%%%%%%%%%%%%%%%%%%%%%%%%%%%%%%%%%%%%%%%%%%%%%%%%%%%%%%%%%%%%%%%%%%%%%%%%%%%%%%%%%%%%%%%%%%%%%%%

\section{X-ray mode switching}
\label{mode_switching}

We have sorted the events collected in the radio Q- and B-mode time windows, respectively, as determined 
in \S~\ref{modes} using our simultaneous radio observations. The accumulated lifetimes in the two radio modes of 
the MOS~1, MOS~2 and Pn cameras
are listed in Table~\ref{lifetimes}. For both modes we applied the two-dimensional 
ML method in the 0.4-1.4 keV band on the pulsar position using the Pn and MOS~1\&2 data and determined the 
total pulsar count rates. It turned out that there is no significant difference in the total X-ray count rates of the Q and B modes, namely 
$0.0132 \pm 0.0005$ and $0.0126 \pm 0.0006$, respectively. 
In order to check whether there is an indication for mode switching in the fluxes of the pulsed and unpulsed 
components, 
in the X-ray  pulse shape, we constructed for the Q- and B-mode data pulse profiles of \psr\ applying 
phase-resolved spatial analysis, thus modelling out the celestial background. 
The pulse profiles (0.4-1.4 keV) are shown in Fig. \ref{Modeswitch} for the Q mode (solid line) and B mode (broken line). 
Applying the Kolmogorov-Smirnov test gives a probability that the two profiles are drawn from the same 
parent distribution of 97.5\% . The differences between the two profiles are shown in the lower figure. This leads to the conclusion 
that we find no evidence for mode switching in the pulsed emission of \psr\ (pulse shape and flux), nor in the flux of the unpulsed emission.

\begin{table}
\caption{ {\bf \psr\ count rates ($\times 10^{2}$) in the energy band 0.4--1.4 keV for different selections on mode lengths in the Q and B modes.}}
\label{modelength}
\begin{center}
\begin{tabular}{c c c c c}
\hline
\hline
Mode length      & Total & 10--210 s & $\ge$ 210 s & \\
\vspace{-2.5mm}\\
  Q mode         &  $1.32 \pm 0.05$ & $1.41 \pm 0.13$ & $1.29\pm 0.05$      &        \\
\hline
Mode length    & Total & 0--50 s &  50 --160 s & $\ge$ 160 s    \\
\vspace{-2.5mm}\\
  B mode      &  $1.26 \pm 0.06$ & $1.59 \pm 0.35$ & $1.20  \pm 0.13$ &  $1.24 \pm 0.07$   \\
\vspace{-2.5mm}\\
\hline \\
\end{tabular}
\end{center}
\end{table}

In Fig.~\ref{Q_B_modewindows} we showed that the mode-length distributions of the Q and B modes are very different. They exhibit
different structures that suggest a relation between mode durations and modulation timescales of the radio intensities
within the modes. Although we do not have an explanation for this intriguing relationship, and do not know whether a relation with the X-ray 
flux can be expected, we made selections on the apparent components in the mode-length distributions to see whether we find variations in
X-ray flux/count rate. 
In the Q-mode-length distribution the initial spike below 10-s duration does not contain a sufficient number of counts, therefore we selected
events detected in the narrow peak at $\sim$50 s between 10 s and the minimum at $\sim$ 210 s ($\sim 6 \times P_3$), and events in 
the trailing tail above 210 s. In the B-mode distribution we selected events in the initial spike until the minimum at $\sim$50 s ($\sim 70P$), the next
component between 50 s and 160 s ($\sim 3 \times  P_3$), and finally all events detected in mode intervals with lengths longer than 160 s.
The derived X-ray count rates for energies 0.4--1.4 keV are given in Table~\ref{modelength} in comparison with the count rates for the total Q and B modes (no selection
on mode length) . For both modes, 
the count rates in the differential mode-length intervals are within one $\sigma$ of the count rates measured for the total Q and B mode data, respectively.
Therefore, we do not see evidence for a relationship between the X-ray count rate and an underlying modulation time scale of the radio intensities within the modes.

\begin{table*}
\caption{ {\bf Spectral fits with power-law (PL) and black-body (BB) models to the total 
emission spectrum of \psr\ over the energy range 0.2-10 keV using Pn and 
MOS~1\&2 data of all observations listed in Table~\ref{table_xmm_obs}. The flux values are derived in a maximum-likelihood spatial 
analysis of sky maps for differential energy intervals.}}
\label{totalspctab}
\begin{center}

\begin{tabular}{ccccc}
\hline
 Model  & PL & BB &  BB+Pl   &   BB+BB    \\
 Fit par.  & Total & Total & Total & Total   \\
\hline
\hline
\vspace{-0.1cm}\\
$N_H$ $\times 10^{-21}$&$3.77_{-0.23}^{+0.26}$ & $1.05 _{-0.12}^{+0.12}$ & $2.83 _{-0.25}^{+0.31}$ & $2.40_{-0.41}^{+0.43}$  \\
\vspace{-0.1cm}\\
$\alpha_{Pl}$ $\times 10^6$& $11.04_{-1.04}^{+1.06}$  & & $6.05_{-1.70}^{+1.64}$            &    \\
\vspace{-0.1cm}\\
$\Gamma$   &$-6.32_{-0.34}^{+0.30}$ &   &  $-5.10_{-0.67}^{+0.60}$ &                 \\
\vspace{-0.1cm}\\
$F_{Pl}$  $\times 10^{14}$  & $8.19_{-0.22}^{+0.20}$&    & $2.59\pm0.20$             &  \\
\vspace{-0.1cm}\\
$\alpha_{BB_1}$& &$0.016_{-0.002}^{+0.003}$  &   $0.27_{-0.12}^{+0.24}$    & $0.43_{-0.14}^{+0.18}$   \\
\vspace{-0.1cm}\\
$kT_1$   & &   $0.126_{-0.004}^{+0.003}$    & $0.085_{-0.007}^{+0.007}$      & $0.083_{-0.004}^{+0.004} $\\
\vspace{-0.1cm}\\
$R_{BB_1}$    & &   $393_{-25}^{+37}$ & $1616_{-359}^{+718}$ &  $2039_{-365}^{+389} $  \\
\vspace{-0.1cm}\\
$F_{BB_1}$ $\times 10^{14}$   & &   $1.70\pm0.04$      &  $2.27\pm0.25$    & $3.2\pm0.2$ \\
\vspace{-0.1cm}\\
$\alpha_{BB_2}$ $\times 10^4$&   & &    & $10_{-5}^{+12}$     \\
\vspace{-0.1cm}\\
$kT_2$    & &     &   & $0.187_{-0.023}^{+0.026} $  \\
\vspace{-0.1cm}\\
$R_{BB_2}$    & &     &    & $98_{-28}^{+48} $  \\
\vspace{-0.1cm}\\
$F_{BB_2}$ $\times 10^{14}$         &      & &             &       $0.65\pm0.11$     \\
\vspace{-0.1cm}\\
$\chi_{r}^2$ / dof   &  1.42 / 33-3& 1.86 / 33-3&  1.24 / 33-5 &  1.14 / 33-5   \\
\vspace{-0.1cm}\\
\hline
\multicolumn{5}{l}{Units: $N_H$ in cm$^{-2}$; $\alpha_{BB}$ in ph/(cm$^2$s\,(keV)$^3$)} \\ 
\multicolumn{5}{l}{$kT$ in keV; Fluxes are unabsorbed for the 0.5-2 keV band in erg/(cm$^2$s);}\\
\multicolumn{5}{l}{$R_{BB}$ in meters adopting a source distance of 1 kpc; $\alpha_{Pl}$ in ph/(cm$^2$s\,keV) at 1 keV}\\
\end{tabular}
\end{center}
\end{table*}

%*************************************************************************** end Table ********************************

%-----------------------------Figure Start------------------------------
 \begin{figure*} 
 \begin{center}
 \vspace{10mm}
   \includegraphics[height=18cm,width=7cm,angle=90]{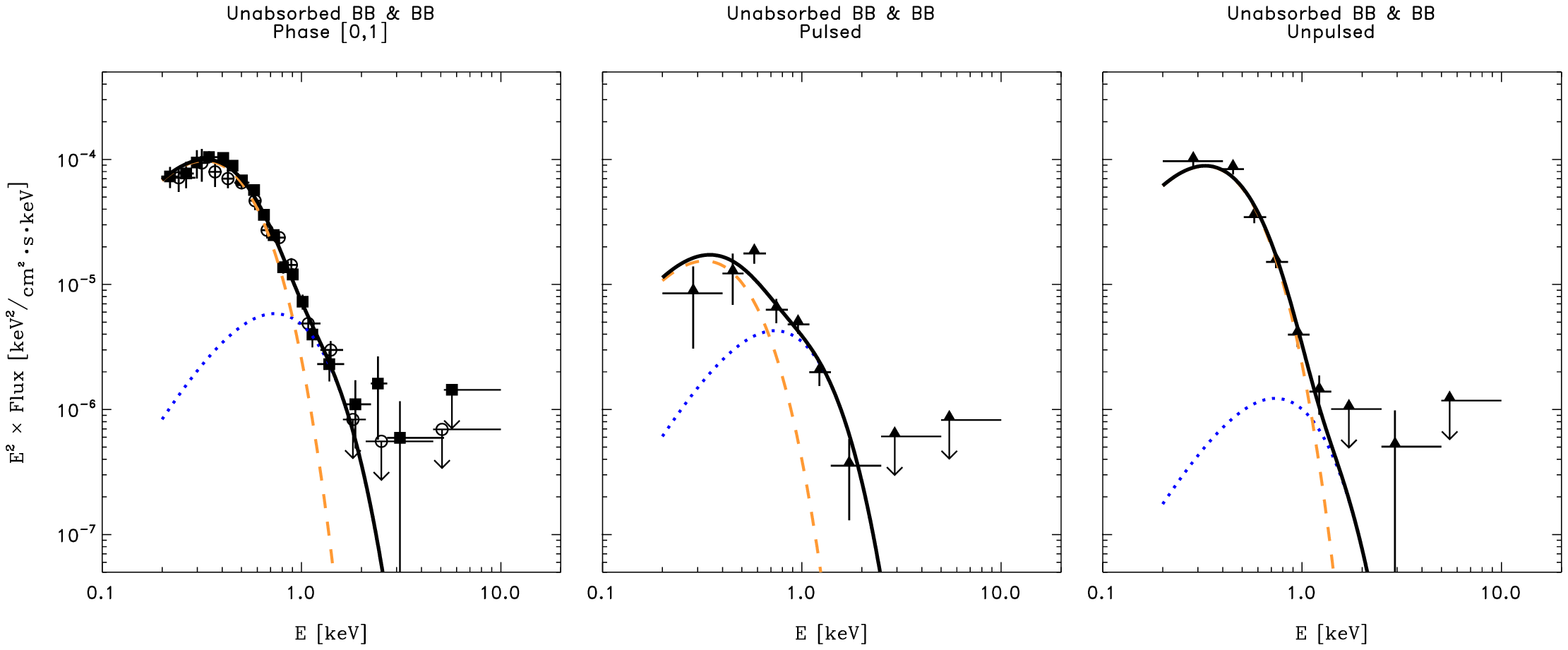}
   \vspace{4mm}
   \caption{\label{PSRB1822spectra} Unabsorbed X-ray photon spectra of \psr . {\it Panel A}: Total spectrum from 
   spatial analyses of count skymaps with flux values independently derived for {\it XMM-Newton} EPIC Pn (filled squares) 
   and MOS~1\&2 CCDs (open circles). 
   The solid line shows the best fit with two BB components to the total spectrum (all phases); the broken orange line shows
   the contribution from the soft component ($kT$ = 0.83 keV) and the dotted blue line the hot component ($kT$ = 1.87 keV). {\it Panel B}: 
   Fits with two BB components with the same $kT$ values to the spectrum of only the pulsed emission, see text.  
   {\it Panel C}: The same for the unpulsed/steady emission. In this case the evidence for the hot component is marginal.
   Error bars are $1\sigma$.} 
  \end{center}
 \end{figure*}
 %-----------------------------Figure End--------------------------------

\section{Spectral analysis}
\label{Spectral analysis}

\subsection{Spectrum total emission of \psr }

We first derived the spectral parameters for the total emission from \psr\ between 0.2 and 10 keV using 
all the observations listed in Table~\ref{table_xmm_obs} the Pn and MOS~1\&2 data and performing the two-dimensional ML analysis. 
We applied the optimisation scheme introduced in Section 4 for the spatial analysis for a grid of user-defined 
energy windows between 0.2 and 10 keV, again adopting for the event pattern $\xi=[0,4]$ and for the flag $\digamma=0$. For energies below 1.4 keV
we fitted with one source at the position of \psr\, and for energies above 1.4 keV we included the second source as well.
This resulted in 
background free source count numbers per energy interval (EPIC Pn and MOS~1\&2 treated separately).
These source counts were converted to source flux values in a forward folding spectral fitting procedure 
assuming a spectral model and using proper response  ({\tt arf} and {\tt rmf} files) and lifetime (dead time 
corrected exposure) information. 

An important parameter in this spectral analysis is the absorbing column density towards \psr, $N_H$. 
An estimate of this parameter is not yet published. Using $HI$ line measurements of pulsars \citet{johnston2001} 
determined an upper limit to the distance of \psr\ of 1.9 kpc, thus a location in front of the Sagittarius-Carina arm. 
We estimated that this would indicate an upper limit 
to $N_H$ of $\sim 3 \times 10^{21} $cm$^{-2}$.  A number of papers quote 
a lower distance, e.g. 0.9 kpc \citep{zou2005, prinz}. Therefore, we first treated $N_H$ as free parameter in our spectral fits 
to verify whether its value from the X-ray spectral analysis will indeed be below the upper limit. 

We first produced fits to the total source-count spectrum with a single power law (${\alpha}E^{\Gamma}$) and a 
single blackbody. Both fits are not acceptable. The power-law index has an unphysical value $-6.32^{+0.30}_{-0.34}$ 
and the single blackbody gave a poor fit with $\chi_{r}^2$ = 1.86 (for 31 degrees of freedom, dof). A fit with a 
blackbody plus power law was also not acceptable with an unphysical value of $\Gamma$ of  $-5.1^{+0.60}_{-0.67}$.
A fit with two blackbodies gave an excellent fit with $\chi_{r}^2 = 1.14 $ (for 28 dof) and a very reasonable value 
for $N_H = 2.40^{+0.43}_{-0.42} \times 10^{21} $cm$^{-2}$.  In conclusion, the total spectrum of \psr\ can successfully be described as 
the sum of two blackbody components, a cooler blackbody with $kT \sim 0.083$ keV (T $\sim 0.96$ MK) and large hot spot radius 
of $\sim 2$ km, and a hotter blackbody with $kT \sim 0.187$ keV (T $\sim 2.2 $ MK) and a radius of only $\sim 100$ m. 
Fig.~\ref{PSRB1822spectra}a shows the excellent fit to the total spectrum with the two blackbodies. Table \ref{totalspctab} lists all spectral 
parameters with their uncertainties of the discussed
fits to the total emission. The best estimate of $N_H$ from this X-ray analysis is
for the two-BB fit ($2.40 \times 10^{21} $cm$^{-2}$).

\subsection{Spectra of the pulsed and unpulsed emissions of \psr }

\label{pulsed_unpulsed_spectra}

In \S~\ref{timing} we introduced the three-dimensional ML approach after we had shown 
that there is no evidence
for pulse shape variations over the {\it XMM-Newton} energy range. In this approach the ML analysis 
is applied in the 3D data space by axes for the defined sky coordinates and pulsar phase. The spatial PSF 
is the point source signature in the skymaps for differential energy intervals and the sinusoid the 
shape of  the pulse profile in all differential energy bins. The complementary pulsed and unpulsed 
source counts per differential energy interval are simultaneously determined. Similar to what has 
been done for the total spectrum, these source counts were converted to source flux values in 
a forward folding spectral fitting procedure 
assuming a spectral model and using proper response  ({\tt arf} and {\tt rmf} files) and lifetime (dead time 
corrected exposure) information. 

In order to compare the fit parameters independent of variations in $N_H$, we fixed $N_H$ in the following fits of the pulsed and 
unpulsed emission to the value obtained for the total emission
with the double BB: $2.40 \times 10^{-21}$ cm$^{-2}$. 
First we fitted the pulsed spectrum with a single power law, and could reject this shape (index $\Gamma$  = -4.3 with 
$\chi_{r}^2$  = 1.87 for 7 dof). However, a single BB rendered an acceptable fit ($\chi_{r}^2$  = 1.37 for 7 dof) with  
$kT = 0.136_{-0.009}^{+0.010}$ keV and hot spot radius R = $256_{-47}^{+73}$ m. The latter two values are in 
between those of the hot and cool 
components in the two-BB fit to the total emission (see $2^{nd}$ column of Table~\ref{pulsed_unpulsedspctab}).
When we then made a fit with a double-BB to the pulsed emission, we find within the statistical significance the same kT
values for a hot and a cool component ($4^{th}$ column)  as 
for the total emission ($2^{nd}$ column). Although statistically not required, the double-BB fit thus renders a consistent picture 
with the same spectral components as obtained for the total emission. For further comparison,  we also fixed 
the two kT values to the best-fit values 
of the total emission (see the $5^{th}$ column of Table~\ref{pulsed_unpulsedspctab}). This interestingly shows 
that the flux of the hot component of the pulsed emission, with small radius of $84\pm 5$ m, is consistent with being
entirely responsible for the flux of the hot component in the total emission. 
The cool component of the pulsed emission contributes 
only a small fraction $(14 \pm 3)\%$ to the flux of the cool component of the total emission. 
This indicates that the cool component (kT= 0.083 keV) should primarily be found in the unpulsed/steady component.

For the spectrum of the unpulsed emission we could immediately rule out a power-law shape
($\Gamma = -5.84$). A single-BB model gave a poor fit ($\chi_{r}^2$  = 2.06 for 7 dof, or a 5\% probability for the fit to be acceptable), 
but with, interestingly, the same cool $kT_{1}$ value of $0.083 \pm 0.003$ keV and hot-spot radius of $\sim$ 2 km as found 
for the cool component in the total fit. 
To see whether there is also
room for the above derived hot component, we added $kT_{2}$ fixed at 0.187 keV. The fit improved
($\chi_{r}^2$  = 1.81 for 7 dof), but there is only marginal evidence for this hot component; the flux has a 50\% error. 
The spectral fit parameters for the unpulsed-emission spectrum are also listed in Table~\ref{pulsed_unpulsedspctab}.

Fig.~\ref{PSRB1822spectra}b\&c show the derived unabsorbed spectra of the pulsed and unpulsed emission, respectively. 
The best fits with two BB components (fixed at kT = 0.083 keV and kT = 0.187 keV) are shown, as well as the individual contributions
of the cool and hot components. We obtained a consistent picture from this spectral 
analysis with the hot (kT = 0.187 keV) BB component in the total emission being due to the events in the pulse at the position of
the main pulse above the flat/unpulsed level. There is not more than a hint for this hot component to be present in the 
unpulsed emission. Furthermore, the cool BB component with kT = 0.083 keV and radius of $\sim$ 2000 m is 
almost entirely due to the unpulsed emission. In this case, such a cool component is also present in the pulsed emission but with smaller radius
of $\sim$ 800 m  and contributes only a small fraction of $\sim14 \%$ to the cool component in the total emission.

%*********************************************************************************************************************************

\begin{table*}
\caption{ {\bf Spectral parameters for single or double black-body (BB) model fits to pulsed and unpulsed emission 
from \psr\ compared to the  fit parameters for the double-BB fit to the phase integrated  [0,1] total emission 
($2^{nd}$ column), the same as the last column in Table~\ref{totalspctab};
 $3^{rd} - 5^{th}$ columns, for only the pulsed emission; $6^{th}, 7^{th}$ columns, for the unpulsed emission. 
The complementary pulsed and unpulsed emission spectra are simultaneously determined in a 
Maximum Likelihood analysis in the 3D data space with axes the sky coordinates and pulsar phase (see text).
Bold values indicate the $N_H$ and $kT$ values fixed to the best-fit values for the phase-integrated 
total emission. Fits are made over the energy range 0.2-10 keV using Pn and MOS~1\&2 data of the observations listed in Table~\ref{table_xmm_obs}. 
}}
\label{pulsed_unpulsedspctab}
\begin{center}

\begin{tabular}{ccccccc}
\hline
Model    & BB+BB &  BB  &  BB+BB   &   BB+BB & BB & BB+BB \\
              &Total  &  Pulsed & Pulsed & Pulsed & Unpulsed & Unpulsed\\
\hline
\hline
\vspace{-0.1cm}\\
$N_H$ $\times 10^{-21}$& $2.40_{-0.41}^{+0.43}$  &  \bf 2.40  & \bf 2.40 & \bf 2.40 & \bf 2.40 & \bf 2.40 \\
\vspace{-0.1cm}\\
$\alpha_{BB_1}$&  $0.43_{-0.14}^{+0.18}$  &  &   $0.034_{-0.026}^{+0.164}$  & $0.068_{-0.014}^{+0.015} $ & $0.39_{-0.10}^{+0.13}$ &  $0.39\pm 0.02$ \\
\vspace{-0.1cm}\\
$kT_1$         & $0.083\pm0.004 $ &  & $0.095_{-0.021}^{+0.025}$      & \bf 0.083  & $0.083\pm 0.003$  & \bf 0.083   \\
\vspace{-0.1cm}\\
$R_{BB_1}$         &   $2039\pm 380$     &  &   $573_{-219}^{+1382}$&  $ 811\pm 89$  & $ 1940\pm 290$  &  $ 1940\pm 290$  \\
\vspace{-0.1cm}\\
$F_{BB_1}$ $\times 10^{14}$         &   $3.2\pm 0.2$  &   & $ 0.75\pm 0.22$   & $0.46_{-0.09}^{+0.10}$  & $2.70\pm0.14$  &  $2.64_{-0.13}^{+0.15}$  \\
\vspace{-0.1cm}\\
$\alpha_{BB_2}$ $\times 10^4$ &    $10_{-5}^{+12}$    &    $6.8_{-25}^{+39}$     &  $7.2_{-5.3}^{+19}$      & $7.3_{-1.0}^{+0.9}$  &  &  $2.1\pm 1.0$ \\
\vspace{-0.1cm}\\
$kT_2$         & $0.187_{-0.023}^{+0.026} $ &    $0.136_{-0.009}^{+0.010} $ & $0.181 _{-0.035}^{+0.050} $    & \bf 0.187 &  & \bf 0.187 \\
\vspace{-0.1cm}\\
$R_{BB_2}$         &    $98_{-25}^{+60}$    &    $256_{-47}^{+73} $ &  $83_{-31}^{+110} $  &  $84\pm 5$   &    & $45\pm 11$   \\
\vspace{-0.1cm}\\
$F_{BB_2}$ $\times 10^{14}$         &     $0.65\pm0.11$  &  $1.12\pm 0.11$    & $ 0.44\pm 0.15$  &  $0.62_{-0.08}^{+0.08}$   &    &  $0.18\pm0.09$   \\
\vspace{-0.1cm}\\
$\chi_{r}^2$ / dof   & 1.14 / 33-5  &    1.37/ 9-2 & 1.44 / 9-4 & 1.42 / 9-2 & 2.06 / 9-2&  1.81 / 9-2\\
\vspace{-0.1cm}\\
\hline
\multicolumn{7}{l}{Units: $N_H$ in cm$^{-2}$;  $kT$ in keV; $\alpha_{BB}$ in ph/(cm$^2$s\,keV$^3$); $R_{BB}$ in meter adopting a source distance of 1 kpc;}\\
\multicolumn{7}{l}{Fluxes are unabsorbed for the 0.5-2 keV band in erg/(cm$^2$s)}\\
\end{tabular}
\end{center}
\end{table*}

%**********************************************************************************************************************

\section{Summary}

Thanks to the long dead-time-corrected exposure of $\sim$ 200 ks with {\it XMM-Newton}, 
together with the unprecedentedly long simultaneous WSRT radio coverage of \psr\, 
we could study its properties in great detail. We first present a summary of the new findings of our
X-ray radio campaign.

	$\bullet$  In $\sim$ 198 ks of useful radio observations 952 mode switches of \psr\ were detected with 63.9\% of the pulses in the Q mode.
	The average Q-mode length was 270 s ($\approx$347 pulses), and average B-mode length 150 s ($\approx$195 pulses). 

	$\bullet$  The histograms of the radio Q-mode and B-mode lengths exhibit very different structures with maxima in one roughly coinciding with minima in the other,
	located at multiples of the modulation periodicity $P_3$. For the first time in a mode-switching pulsar this provides evidence for a relationship between the duration of its 
	modes and a known underlying modulation timescale of the radio intensities within the modes.

	$\bullet$  In the X-ray spatial analysis \psr\  was detected together with a nearby hard-spectrum X-ray source located at only $(5.1\pm 0.5)\arcsec$ from the X-ray position of
	\psr\ ($\alpha_{2000} = 18^{\hbox{\scriptsize h}} 25^{\hbox{\scriptsize m}} 30\fs726 , \delta_{2000} = -9\degr 35\arcmin 23\farcs14$).	 The new source dominates over \psr\
	in skymaps for energies above 1.4 keV, and might be a PWN. 
	
	$\bullet$ The observations revealed X-ray pulsations from \psr\  between 0.4 and 1.4 keV with a broad sinusoidal pulse detected at a significance 
	of $9.6 \sigma$. The X-ray profile reaches its maximum at phase $0.094 \pm 0.017$, slightly lagging the radio main pulse that peaks at phase 0, and is shifted in the opposite direction
	w.r.t. the position of the radio precursor at phase --0.042 (or 0.958). 
	
	$\bullet$ The X-ray phase distribution does not show an X-ray pulse at the phase of the radio interpulse, shifted by 180\degr w.r.t. the phase of the main pulse. 
	If \psr\ emits an X-ray pulse at the phase 
	of the interpulse, weaker than the X-ray pulse at the phase of the radio main pulse, but with similar sinusoidal shape, then we do not expect 
	to see this pulse above an unpulsed level that is created by the sum of the X-ray interpulse and the same flux from the X-ray main pulse.

	$\bullet$ The pulsed fraction varies with energy from $\sim$~0.15 around 0.3 keV to a value of $\sim$~0.6 around 1 keV, indicative for a significant 
	spectral difference between the pulsed and unpulsed emissions. 
	
	$\bullet$ The observations did not reveal evidence for simultaneous mode switching in the X-ray and radio bands, nor in the X-ray pulsed emission 
	(pulse shape or flux), nor in the flux of the unpulsed X-ray emission. The total count rate for energies 0.4 -- 1.4 keV in the Q mode, $(1.32 \pm 0.05) \times 10^{-2}$, equals
	the count rate in the B mode, $(1.26 \pm 0.06) \times 10^{-2}$.	
	
	$\bullet$ There is no evidence for a relationship between the X-ray count rate and the underlying modulation time scale $P_3$ of the radio intensities within the modes.	
	X-ray count rates for apparent different structures/components in the histograms of the radio Q-mode and B-mode lengths rendered the same count rates as obtained
	for the total Q and B mode.
	
	$\bullet$ The total X-ray spectrum is well described by a double blackbody model with a cool component with $kT = 0.083\pm 0.004$ keV or 
	T =$0.96\pm 0.05$ MK and a hot component with $kT = 0.187\pm 0.026$ keV or T =$2.17\pm 0.30$ MK.  The first has a hot-spot radius of 
	$2.0\pm 0.4$ km and the latter $98_{-28}^{+48}$m. 	
	
	$\bullet$ The spectrum of the pulsed emission is well fit with a single blackbody, or a double blackbody with temperatures which are consistent with the values 
	obtained for the hot and cool components in the spectrum of the total emission. For the latter fit, the flux of the hot component ($kT \sim 0.187$ keV with 
	hot-spot radius of $\sim$84 m) is consistent with the flux of the same component in the total-emission spectrum.
		
	$\bullet$ The spectrum of the complementary unpulsed emission	is almost entirely explained by the cool BB component with kT = 0.083 keV and 
	hot-spot radius of $\sim$ 2000 m. This fit can be somewhat improved by adding a small contribution of the hot component identified in the total spectrum.

 \section{Discussion}
 
 To date, long simultaneous X-ray and radio campaigns on mode-switching pulsars have only been performed on PSR B0943+10
 and \psr. For PSR B0943+10, \citet{hermsen2013} discovered simultaneous mode switching in X-ray and radio emissions,
 not seen in our campaign on \psr. What causes the simultaneous mode switching is still an enigma. For PSR B0943+10, and now also
 for \psr,  the pulsed X-ray signals were detected for the first time, and the X-ray pulse profiles are similar; both exhibit a
 single sinusoidal pulse on top of unpulsed emission from the pulsars. This seems remarkable, because PSR B0943+10
 is reported to be a nearly-aligned rotator with our line of sight passing near the pole \citep{deshpande2001}. Therefore, we continuously  
 view isotropic thermal emission from one single pole of this pulsar, and no pulsation of thermal emission is expected.
 For \psr\  we conclude in the Appendix that the existing evidence of
the known radio characteristics of \psr\ strongly suggests this pulsar as having an orthogonal rotator geometry with our line of sight close to the equator.  
In this case we view such thermal emission from hot spots on both poles every rotation.
Further in the discussion we will address similarities of, and differences between these two pulsars where appropriate.

 \subsection{The nature of the radio modes}
 
 The contrast between the two radio modes of \psr\ could hardly be greater, going well beyond their original nomenclature of B (``bright'') and Q (``quiet'): 
 in Fig.~\ref{WSRT-modes-PSR1822-09} the precursor (the best mode indicator and used here to define the mode) can be seen to sharply switch on and off in tandem with the intensity of the 
 main pulse and in anti-correlation with the interpulse strength. Given the clear switch in X-ray emission found between the similarly-named 
 modes of PSR B0943+10, one might expect the same effect in \psr. 

However, we report here that no radio vs X-ray correlation has been found (a 15\% change in X-ray count rate could have been detected
at an $\approx 3\sigma$ level), and it is interesting to consider how \psr\ differs from PSR B0943+10. 
The most obvious difference is that the magnetic axis of \psr\ is almost certainly highly inclined to its rotational axis (see Appendix), 
while PSR B0943+10 is in near alignment. Another difference lies in the nature of the modal modulations found in both pulsars. In PSR B0943+10 the 
B-mode exhibits a highly regular and rapid ($\sim2P) P_3$ periodicity which may last for several hours, while in the Q-mode this 
modulation vanishes and is replaced by highly disordered emission. The conventional interpretation of this is that a precisely circulating 
carousel on the polar cap dissolves and is replaced by chaotic discharging until the order is suddenly recovered. Thus the mode lengths 
far exceed the subpulse modulation timescale.

In contrast, \psr\ has been found to have modulation in both modes, both on a much longer timescale ($\sim46.6P$ and $70P$) than that of 
PSR B0943+10 and, significantly, not showing any subpulse ``drift'' found in carousel-driven modulation (not for orthogonal rotators though). The modulated emission is 
diffuse against a steady background (and often hard to discern).  This suggests a different physical mechanism and the fact that both 
the mode change and the modulations occur simultaneously at both poles suggests that both reflect magnetospheric effects, rather than local polar cap physics. 

In light of this it is perhaps not surprising that we find a link between mode lengths and modulation periods in \psr. In Fig.~\ref{Q_B_modewindows}(upper) 
we see that the most likely Q-mode length is just the Q-mode modulation period ($46.55P$) and in Fig.~\ref{Q_B_modewindows}(lower) that the most likely B-mode 
lengths are double and quadruple multiples of the B-mode modulation ($70P$). The further fact that the modulation periods of the 
two modes are harmonically related and that the modes probably (though this is unproven) start at the same phase of their modulations 
(either peak or trough) suggests that the modes may both be manifestations of a common underlying magnetospheric ``clock'', 
possibly set by the inclination of the pulsar, the size of its light cylinder and its magnetic field strength. 

This view has the advantage of divorcing the radio emission from polar cap conditions and suggesting a commonality to both 
modes, so that the uncorrelated thermal X-ray emission may indeed relate to an unchanging polar cap temperature. It is 
interesting to note that the two radio-detected pulsars of the ``Three Musketeers''\footnote{\citet{becker1997} dubbed PSR B0656+14, PSR B1055$-$52
and Geminga / PSR J0633+1746 the Three Musketeers. These three middle-aged pulsars have similar characteristics and were the only 
pulsars in the first list of 27 X-ray detected pulsars that exhibited thermal X-ray components (then probably also by the Vela pulsar), which could be attributed
to thermal emission from the neutron star stellar surface; so called cooling neutron stars.} (PSR B0656+14 and PSR B1055$-$52) exhibit 
non-drifting modulations with timescales of $20P$. In the case of PSR B0656+14,  Weltevrede et al. (2006, 2012) identify two distinct kinds of 
polar cap radio emission, of which only the ``spiky'' emission is modulated. It would be of interest to investigate this in further 
single-pulse studies of \psr. The modulated and unmodulated radio emission could be identified with the two 
distinct pulsed blackbody components already inferred for the X-ray emissions of PSR B0656+14, PSR B1055-52 and Geminga by \citet{deluca2005} and proposed here for \psr.    

 \subsection{\psr, a middle-aged pulsar like the Three Musketeers?}

The total emission spectrum of \psr\ was well described by the sum of a hot BB and a cool BB spectrum.
This differs from our findings for PSR B0943+10. In that case the total emission spectrum appeared to be
the sum of a hot BB and a power-law spectrum. PSR B0943+10 has a long spin period of 1.1 s and is relatively old (5 Myr)
compared to \psr\, which has a shorter period of 0.77 s and is younger (233 kyr). Also, the spin-down luminosity of \psr\
is a factor $\sim$40 ($4.6\times 10^{33}$ erg s$^{-1}$) larger than that of PSR B0943+10. The characteristics of \psr\ might better be 
compared with those of the Three Musketeers. These pulsars have similar 
characteristic ages, but rotate a factor 2-4 faster with spin-down luminosities up to a factor $\sim$8 larger than \psr. 
The Three Musketeers exhibit in the 0.2-8 keV band spectra with three components: a cool BB (T=0.50-0.79 MK), a hot BB (T=1.25-1.90 MK)
plus a PL component that contributes negligibly below $\sim$2 keV and dominates above (see \citet{deluca2005}). The two BB components of \psr\
have temperatures only slightly higher than those of the Three Musketeers and the fluxes of their PL components are too low for detection when scaled to
the an-order-of-magnitude-lower count rate from \psr.

%%%%%%%%%%%%%%%%%%%%%%%%%%%%%%%%%%%%%%%%%%%%%%%%%%%%%%%%%%

\begin{figure}
 \begin{center}
     \includegraphics[width=82.5mm]{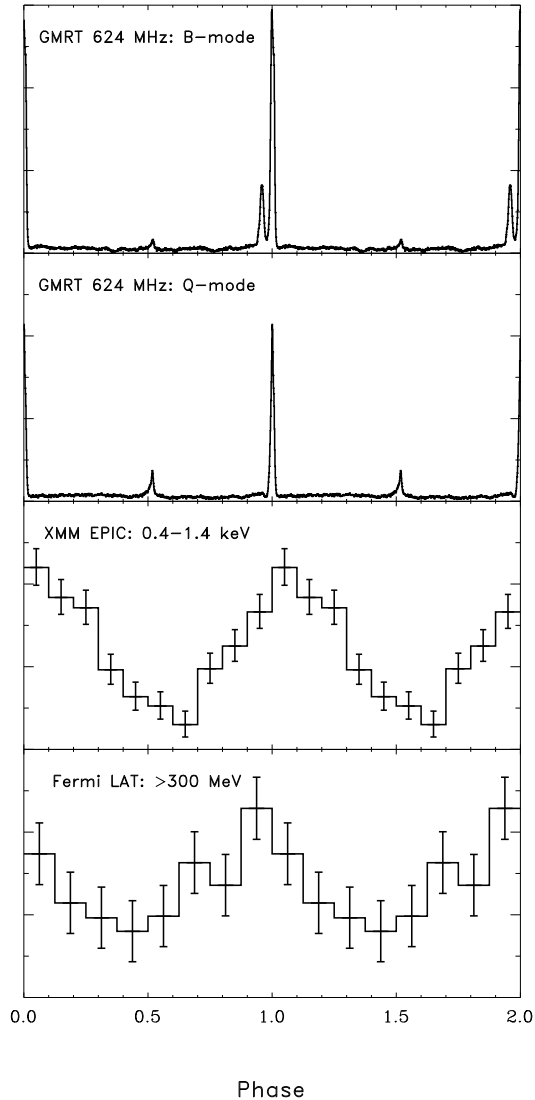}     
     \caption{\label{B1822Fermi} Pulse profiles of \psr, from top to bottom: GMRT radio profiles at 624 MHz in arbitrary units for the B and Q mode, respectively; {\it XMM-Newton} EPIC X-ray profile 
     for energies 0.4 to 1.4 keV; {\it Fermi} LAT profile for energies above 300 MeV. The vertical axes of the X-ray and gamma-ray pulse profiles are 
     in counts/bin, and the error bars are statistical only (1 $\sigma$).}
  \end{center}
\end{figure}

%%%%%%%%%%%%%%%%%%%%%%%%%%%%%%%%%%%%%%%%%%%%%%%%%%%%%%%%%%%

The Three Musketeers have also been detected at high-energy gamma rays above
100 MeV, requiring a very hard steepening PL spectrum above the X-ray band in order to bridge several decades in 
luminosity to the high-energy gamma-ray band. 
Given the resemblance with \psr\ in the X-ray band, we selected all {\it Fermi}-LAT Pass 8-event
 data collected between August 4, 2008 and
 December 9, 2015 (MJD 54682.897 -- 57365.907) with \psr\ in the field-of-view (region-of-interest: circular aperture of radius $5\degr$ centered on the pulsar), and
 extended the Jodrell Bank ephemeris to cover this time window.
 Using {\it Fermi} Science Tools\footnote{http://fermi.gsfc.nasa.gov/ssc/data/analysis/software/} 
 gtselect and gtbary, respectively, we further selected only events with Evclass = 128, evtype=3 and Earth zenit angle $<105\degr$, and
 barycentered  the event arrival times using the Solar System Ephemeris file DE200.  
 We selected events with energies above 300 MeV, 
 because we expect rather hard gamma-ray emission from
 middle-aged pulsars.  Namely, \citet{posselt2015} list the 11 nearby middle-aged pulsars detected by the LAT. They have 
 power-law spectra with average index of $\sim$ $-$1.1 and average exponential cutoff of $\sim$ 1.3 GeV. 
 We phase folded the barycentered arrival times of events
 that were detected from within an
 energy-dependent aperture around \psr\ that contains $\sim$68\% of the point-source counts ($\theta_{68\%}$ [degree] = $0.8 \times$ (E[MeV]/1000)$^{-0.75}$) \citep{abdo2009}.
 We verified that for the energy range where we might expect to detect events from \psr\ (up to a few GeV) 
 this parametrisation is consistent with the most recent one derived for Pass 8-event 
 data{\footnote{http://www.slac.stanford.edu/exp/glast/groups/canda/lat\_Performance.htm; http://fermi.gsfc.nasa.gov/ssc/data/analysis/documentation/Cicerone/Cicerone
 \_LAT\_IRFs/IRF\_PSF.html}.} 
 The obtained pulse profile is shown in the bottom panel of Fig. \ref{B1822Fermi}. There is an indication for the detection of a pulsed signal at a significance level of
 $\sim 2.7 \sigma$ \citep[][$Z_1^2$ value]{buccheri1983}, or a $0.66\%$ chance probability.
Fig.~\ref{B1822Fermi}  shows the gamma-ray profile in comparison with those in the radio band for the Q and B mode 
and the X-ray profile. It is interesting to note that the maximum of the broad gamma-ray profile at phase 
0.92 $\pm$ 0.05 approximately coincides in phase with the PC in the radio B mode, and is shifted by 3.3 $\sigma$ with respect 
to the maximum of the X-ray pulse at phase 1.094 $\pm$ 0.017. We note, that the {\it Fermi}-LAT Third Source Catalog \citep{acero2015} does not report a 
source within a distance of a degree of \psr.

\subsection{Modelling of the pulsed fraction}

The apparently (nearly) orthogonal rotator geometry of \psr\ seems difficult to reconcile with the energy dependence of the pulsed fraction, 
reaching a maximum value of $\sim$ 0.6 at 1 keV, for a single hot (T =$2.17\pm 0.30$ MK) pulse
on top of cool (T =$0.96\pm 0.05$ MK) unpulsed emission. In fact, we derived from the spectral analysis that the hot pulsed component has
a pulsed fraction that could be as high as 100\%. 

For an orthogonal-rotator geometry it is remarkable that we did not see evidence for an X-ray pulse in the pulse profile
from the phase of the IP. However, as we discussed, such a pulse can be hidden in the pulse profile in a flat
unpulsed emission level when the X-ray MP and IP both are sinusoidal. The fit with two BBs to the spectrum of the unpulsed emission 
gives a hint for the presence of a weak hot component. If this hot component in the unpulsed emission is genuine, then in first approximation 
$\sim$ 50\% can be assigned to the MP and $\sim$ 50\% to the IP
(assuming both have a sinusoidal shape).  The best-fit flux values then suggest
an IP flux of $\sim$12\% of that of the MP and a pulsed fraction of just the hot component of $\sim$ 0.8.
These high pulsed fractions cannot be produced by isotropic blackbody emission in hot spots with the same temperatures and areas on
both poles for the geometry assigned to \psr. When we take the approximate relation ${F_{BB} \propto T^4}$ then 
the temperature of the antipodal spot (for the same area)
should by $\sim 59\%$ of that of the hot spot producing the main X-ray pulse to reach a pulsed fraction of $\sim$ 0.8. 
This would mean a temperature of $\sim $0.110 keV.

For PSR B0943+10 we obtained similarly high pulsed fractions \citep{hermsen2013},
namely, the pulsed fraction in the Q mode of a single hot thermal pulse on top of non-thermal unpulsed emission
increased from $\sim$ 0.2 at 0.2 keV up to $\sim$ 0.6 at 1 keV. Considering just the thermal pulsed component in the Q mode,
the modulation was also reported to be consistent with 100$\%$. These pulsed fractions are confirmed in the longer follow-up
campaign on this pulsar \citep{mereghetti2016}. 
PSR B0943+10 has an estimated magnetic field strength at the pole of  $ 2 \times 10^{12} $ G, significantly above 
a field strength of $ B = 10^{11} (E/1 keV) $ G, when the electron cyclotron energy $E_c$ exceeds the photon energy,
and magnetic beaming makes the local emission essentially anisotropic (beamed along the direction of the magnetic field). 
The properties of a neutron star atmosphere (e.g. fully or partially ionised) determine the angular distribution and shape of the 
pulsations. The geometry and viewing angles determine finally what we observe. For example, \citet{pavlov1994} showed that for a fully 
ionised neutron star atmosphere a strong narrow pencil beam along the magnetic field direction can be expected 
together with a broad fanlike component across the magnetic field. Even in the case of a uniformly heated neutron 
star surface the angular distribution is beamed along the magnetic axis \citep{zavlin2002}. 
More recently, \citet{adelsberg2006} also considered atmosphere models of magnetized neutron stars, and calculated 
characteristic beaming patterns exhibiting a thin pencil beam at low emission angles and a broad fan beam at large 
emission angles. See also similar work by \citet{zane2006} and \citet{turolla2013}.

Stimulated by the discovery of simultaneous X-ray radio mode switching \citep{hermsen2013}, 
\citet{storch2014} considered a magnetized partially ionized hydrogen atmosphere model
to explain the high pulsed fractions in the Q mode of PSR B0943+10, while remaining consistent with the nearly
aligned dipole field geometry  \citep{deshpande2001}, and using the magnetic field strength of $ 2 \times 10^{12}$ G inferred from its $P$ and $\dot{P}$. 
They derived pulsed fractions ranging from 0.4 at 0.2 keV, up to a maximum of 0.9 at a few keV, that are higher than the pulsed fractions of the total emission.
However, their calculations assumed just a single standard polar cap hot spot, and do compare well with, or are somewhat lower than,
the measured values for the total emission when the calculated pulsed fractions are corrected (lowered) for the underlying unpulsed emission \citep[see also][]{mereghetti2016}.

\psr\ has an even stronger estimated magnetic field strength of $6.4 \times 10^{12}$ G such that also for this pulsar magnetic beaming
can shape the measured pulse profile. For its  orthogonal geometry,
we obviously need strong beaming effects such that hot X-ray emission from the IP is totally or mostly beamed away from
our line-of-sight, while the X-ray emission from the MP passes through our line-of-sight. 
The cool unpulsed component in the total emission from \psr, originating from an area with large radius of $\sim$ 2 km, can be 
explained as the sum of broad pulses from both poles, possibly in fan beams. This would explain the finding
that the estimated radius is significantly smaller than one expects for cool emission originating on the entire surface of
the neutron star. This is in contrast to the cool areas found for e.g. the Three Musketeers, that are (more than) an order of magnitude larger,
and their cool emission is thought to originate from the entire surface.

Following the above argumentation, if the thermal X-ray emissions from PSR B0943+10 and \psr\ are strongly beamed, and mode switching is due
to some change in structure of the magnetic field, then for \psr\ the beaming angle
of the X-ray pulsations might change less w.r.t. our line of sight between the B and Q mode than in a mode switch of PSR B0943+10. 
The lack of observable change in the pulse shape and brightness  of \psr\ can then just be a coincidence of viewing geometry.

\subsection{Conditions at the inner acceleration region}

It is believed that the hot spots are actual polar caps heated by backstreaming ultra-relativistic particles accelerated in the inner acceleration region.
Thus, the study of properties of hot spot components, such as size and temperature, can provide valuable information about this region of a neutron star.
One can use the magnetic flux conservation law to estimate the magnetic field at the polar cap region $B_{\rm s} = B_{\rm d} R_{\rm dp}^2/R_{\rm BB}^2$.
Here, $B_{\rm d}$ is the dipolar component of the magnetic field at the polar cap, and $R_{\rm dp}$ is the radius given by 
the last open field line of the dipolar magnetic field component.
Old pulsars with their spectrum dominated by hot spot radiation show that an actual polar cap is considerably smaller than $R_{\rm dp}$, see e.g. 
PSR J0108$-$1431 \citep{pavlov2009, posselt2012}, PSR J0633+1746 \citep{kargaltsev2005}, PSR B1929+10 \citep{misanovic2008}.
Furthermore, the estimated magnetic field strengths at the polar cap for these pulsars ($B_{\rm s} \sim 10^{14}$G) and temperatures of a few million Kelvin 
satisfy the condition for the formation of a Partially Screened Gap (PSG) \citep[see][]{gil2003, medin2007}.
Although, the fitted hot spot radius of \psr\ ($R_{BB}\sim 100\,{\rm m}$) is smaller than the conventional polar cap size 
($R_{\rm dp}=165\,{\rm m}$) the derived magnetic field strength at the polar cap $B_{\rm s}=2.7 \times 10^{13} \, {\rm G}$ and 
the fitted temperature $T\approx 2.2 \times 10^{6} \, {\rm K}$ do not satisfy the condition for PSG formation  \citep[see Eq. 5 in][]{szary2015}.
For such a high surface temperature and relatively weak surface magnetic field the outflow of thermal ions from the surface screens the gap completely.
It is worth noting that both the size and temperature of the hot spot depend on the considered model. The atmospheric models, suggested by the high pulsed fraction, 
could result in different properties of the polar cap, and thereby different conditions in the inner acceleration region.
But, an alternative acceleration mechanism might be at work, for instance the space-charged-limited flow model \citep{arons1979, harding2002}. In that case, 
surface charges can freely flow into the magnetosphere and the inner gap is thought to be space-charge limited. Pairs are created higher up in the magnetosphere
near the so-called ``pair-creation front''.  \citet{harding2001} discuss pulsar polar cap heating and surface thermal X-ray emission in the case of Curvature Radiation pair fronts,
and \citet{hardingm2002} in a second paper the case of Inverse Compton Radiation pair fronts. For the age and temperature of \psr\ only the first scenario might explain
our measured thermal X-ray luminosity. 

\section{Conclusions}

We have observed \psr\ for $\approx{55}$ hours simultaneously at X-ray and radio frequencies. The X-ray results reveal a single-peaked, 
thermal sinusoidal pulse combining a small pulsed hot spot together with a 2km-wide cooler region. Had we observed the 
pulsar at X-rays alone we might have concluded it was a middle-aged Geminga-like pulsar at low/intermediate inclination so that only a single polar hot spot is 
visible to us. This result was totally unexpected since it is known that at radio frequencies this pulsar displays an interpulse, and there is evidence from the
polarisation that its magnetic axis is highly inclined to the rotation axis -- so why doesn't the X-ray emission have a double peak? We argued that a 
weak broad pulse could be hiding in the unpulsed emission.

A second puzzle arises from the lack of modulation in the X-ray pulse. For many years it has been known that \psr\ switches in the radio band between 
two very distinctive modes of emission centred on the magnetic polar regions on a timescale of several minutes. During our observations we 
collected the largest ever number of mode transitions, and so for the first time we were able to build an overview of the statistics of the mode switches. 
These suggest a complex relationship between the durations of the modes and independently-reported modulation timescales present within each 
mode. Following the example of PSR B0943+10, where the pulsed and unpulsed X-ray emissions in its two radio modes differ in flux by a factor 
$\sim$ 2 and the polar hot-spot temperature seems to switch between two values \citep{mereghetti2016}, one might expect a detectable modulated response also 
in the X-ray emission of \psr. However, 
despite multiple attempts at matching our X-ray counts with aspects of the radio modulations, we found no correlation between the X-ray emission 
and simultaneously observed radio emission mode, e.g. the total count rates in the two modes are within $\sim 5\%$ ($1 \sigma$) the same.
At face value, it seems that all the X-ray and radio emission of \psr\ have in common is the same rotation 
period and the same phase longitude for the main pulse. 

Interpreting the X-ray observations of \psr\ we conclude that the X-ray emission may consist of emission from a principal small hot spot at the main pulse and two 
broader and cooler hot spots at the main pulse and interpulse phase, the first being $\sim$ 35\% stronger than the latter. 
This may be explained by exploiting the effects of photon beaming in a strong magnetic field, with the hot X-ray 
interpulse not being directed towards us. In this context it may be worth remembering that the radio MP/IP ratio is (also) quite extreme (see Fig.~\ref{B1822Fermi}). 
Nevertheless, the puzzle remains as to why there appears to be no correlation between the thermal 
emission of the X-rays and the dramatic pan-magnetospheric mode changes implied by the radio emission. This outcome therefore forms an 
extreme contrast to the equivalent observations of the likely near-aligned pulsar PSR  B0943+10 \citep{hermsen2013, mereghetti2016}.

One way out of this dilemma may be to argue that while the thermal X-rays come from the neutron star surface, the radio emission 
may originate from magnetospheric effects which do not directly impact the star's polar caps or its wider surface. The polar interplay of the 
B and Q modes gives little doubt that some form of communication exists between the poles, and this must take place within the supposedly 
closed magnetosphere. Furthermore, it can be shown that the PSG (partly screened gap) mechanism for the production of polar cap patterns 
immediately above the surface will probably not work given the weak inferred magnetic field strength of \psr. So an alternative 
mechanism, presumably operating at higher altitudes in the magnetosphere, must be responsible for the radio emission. Nevertheless, we 
still have to explain how the pulsar's apparently unmodulated hot-spot surface temperatures can be maintained at such high levels.

Finally, following the findings of \citet{lyne2010} that \psr\ is one of a number of pulsars whose long-term profile modulations can be 
linked to changes in spin-down rates, it seems we must now come to the unexpected result that this pulsar's thermal X-rays do not,
or hardly, 
participate in the spin-down mechanism.

The results of our campaign on \psr\ have therefore raised more questions than answers. There is a clear disconnect between the 
radio and  X-ray emission of this pulsar, and at times it has felt as though we are observing two different pulsars with the same period. One 
conclusion is certain: we are still far from understanding either the radio or the X-ray production mechanisms of pulsars.

\section*{Acknowledgments}

We thank the staff of {\it XMM-Newton}, WSRT, GMRT and Lovell for making these observations possible. {\it XMM-Newton} is an ESA science mission
with instruments and contributions directly funded by ESA member states and by NASA. The WSRT is operated by the Netherlands Institute for Radio Astronomy (ASTRON).
GMRT is run by the National Centre for Radio Astrophysics of the Tata
Institute of Fundamental Research. Pulsar research at the Jodrell Bank Centre for Astrophysics and the observations using the Lovell telescope is 
supported by a consolidated grant from the STFC in the UK. We thank Christine Jordan and Andrew Lyne for assistance with the data acquisition and 
Cees Bassa for developing the ROACH backend and for assistance with the data analysis. 

ASTRON and SRON are supported financially by the Netherlands Organisation for Scientific Research (NWO).

JMR acknowledges funding from US NSF grant 09-68296 and a NASA Space Grant.
JWTH acknowledges funding from an NWO Vidi fellowship. 
JWTH and JvL acknowledge funding from the European Research Council under the European Union's 
Seventh Framework Programme (FP/2007-2013) / ERC Starting Grant agreement nr. 337062 (``DRAGNET'') and nr. 617199 (``ALERT''), respectively.
A.S. acknowledges support by NWO under project ``CleanMachine'' (614.001.301).

The {\it Fermi} LAT Collaboration acknowledges generous ongoing support
from a number of agencies and institutes that have supported both the
development and the operation of the LAT as well as scientific data analysis.
These include the National Aeronautics and Space Administration and the
Department of Energy in the United States, the Commissariat \`a l'Energie Atomique
and the Centre National de la Recherche Scientifique / Institut National de Physique
Nucl\'eaire et de Physique des Particules in France, the Agenzia Spaziale Italiana
and the Istituto Nazionale di Fisica Nucleare in Italy, the Ministry of Education,
Culture, Sports, Science and Technology (MEXT), High Energy Accelerator Research
Organization (KEK) and Japan Aerospace Exploration Agency (JAXA) in Japan, and
the K.~A.~Wallenberg Foundation, the Swedish Research Council and the
Swedish National Space Board in Sweden.
 
Additional support for {\it Fermi} science analysis during the operations phase is gratefully 
acknowledged from the Istituto Nazionale di Astrofisica in Italy and the Centre 
National d'\'Etudes Spatiales in France.

\appendix
\section{The Geometry of \psr}

PSR B1822$-$09 exhibits two distinct emission modes.  In its `B'right mode it shows a main pulse (MP) 
along with a precursor component (PC), and in its `Q'uiet mode one sees a weaker and more 
complex MP along with an interpulse (IP).  This IP is separated by almost exactly 180$^{\circ}$ 
(see \citep{backus2010}, hereafter BMR10), although strangely the MP and IP show 
correlated intensity fluctuations while in principal they should have independent emission 
characteristics \citep{fmw81, fw82, dyks2005, gil1994}.
It is however this 180$^{\circ}$ separation seen in the Q-mode between the MP and IP which leads 
one to conjecture that the pulsar might be an orthogonal rotator 
\citep[e.g.][hereafter DZG05, BMR10]{hankins1986, dyks2005}.
Although two further checks are necessary to strengthen the orthogonal rotator 
argument:  the first is that the 180$^{\circ}$ separation between the MP and IP is seen 
over the entire band of observable frequencies, and the second is that the linear polarization position 
angle (PPA) traverse should be adequately described by the rotating vector model (RVM).  
Unfortunately no exhaustive study of these effects exists in the literature, possibly due to the 
lack of high-quality single-pulse observations permitting  separation and individual study of the 
B and Q modes\footnote{Note that the recent PhD thesis of Phrudth Jaroenjittichai, 
https://www.escholar.manchester.ac.uk/uk-ac-man-scw:199950, University of Manchester, 
dedicates a chapter to the study of \psr .}. As pointed out by DZG05 the 180$^{\circ}$ MP-IP 
separation can also be explained in almost aligned rotator 
models where the MP and IP are produced by the line of sight cutting through two sides 
of the hollow conical emission beam.  However in this case, due to radius to frequency 
mapping one expects to see the MP-IP separation evolving with frequency.  The other 
alternative in the aligned rotator category is a single-pole model, wherein the line of 
sight always remains within the emission cone and under special situations emission 
components separated by 180$^{\circ}$ are seen. Although in these cases (often known 
as wide profile pulsars) a low level bridge of emission is expected between the profile
components, which however is completely absent in PSR B1822--09.

%***********************Begin Figure************************************************

\begin{figure*}
\begin{center}
%\begin{tabular}{@{}lr@{}}
{\mbox{\includegraphics[width=70mm]{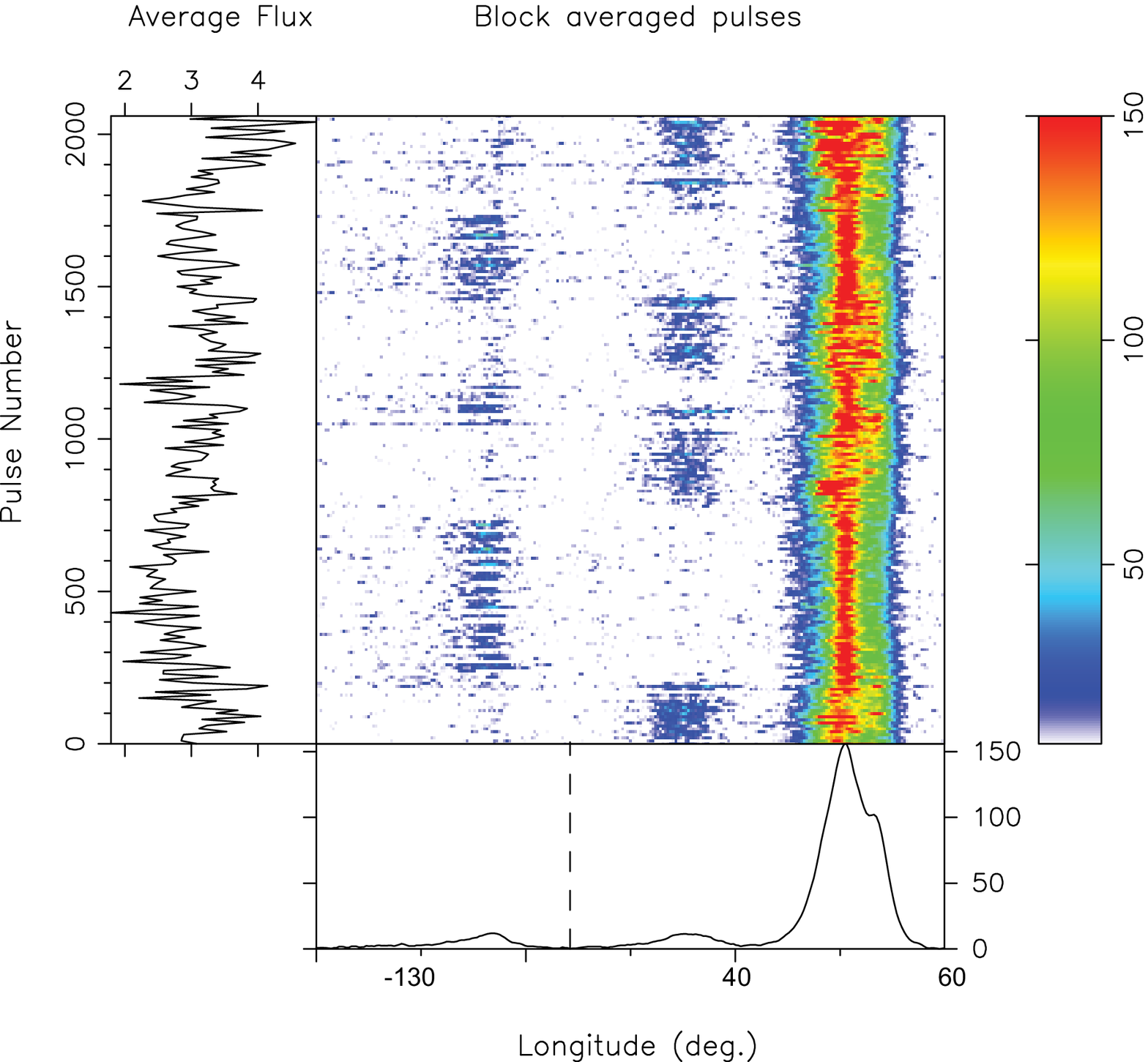}}}
{\mbox{\includegraphics[width=70mm,angle=-90,origin=c]{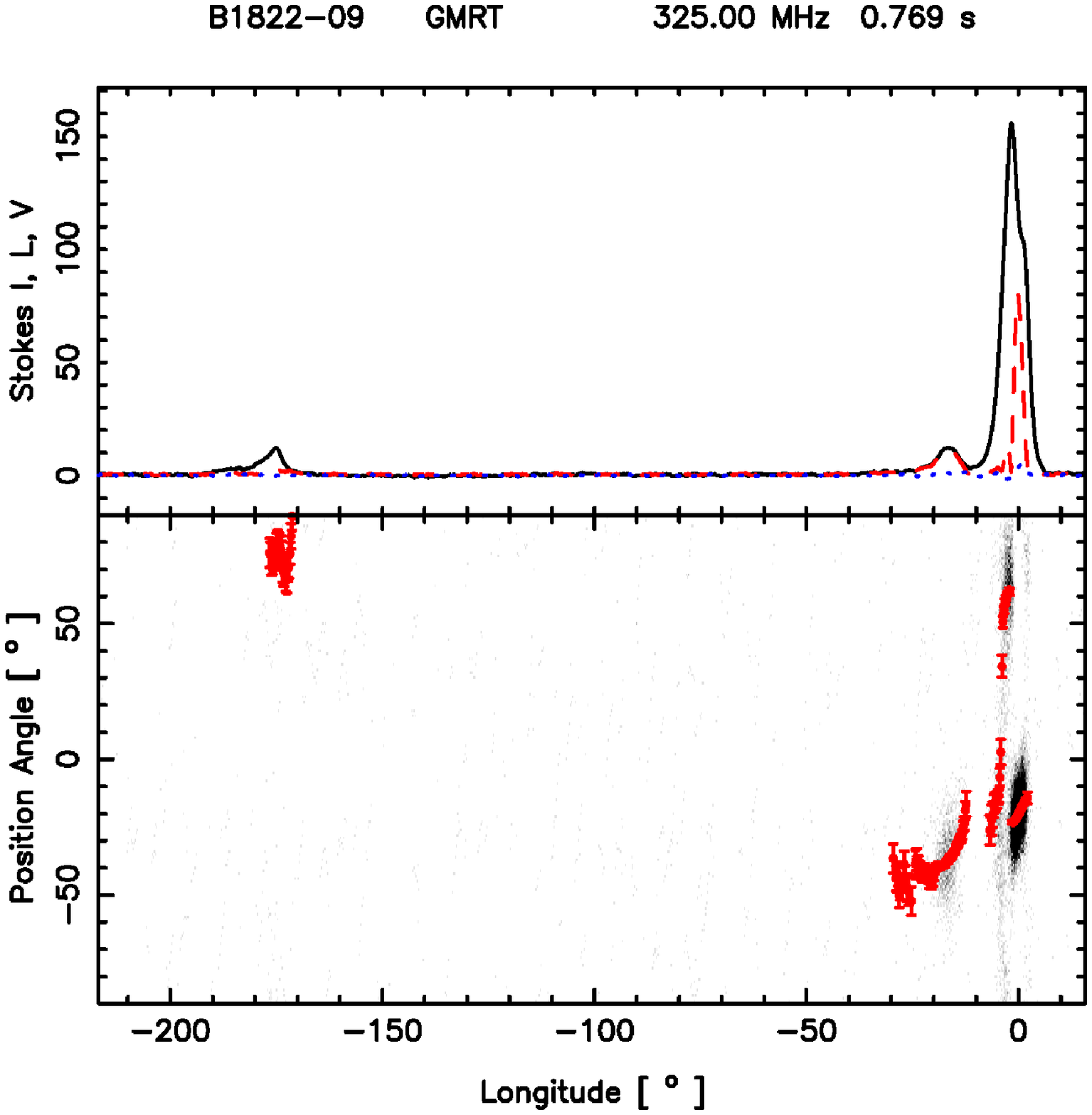}}}
{\mbox{\includegraphics[width=70mm,angle=-90,origin=c]{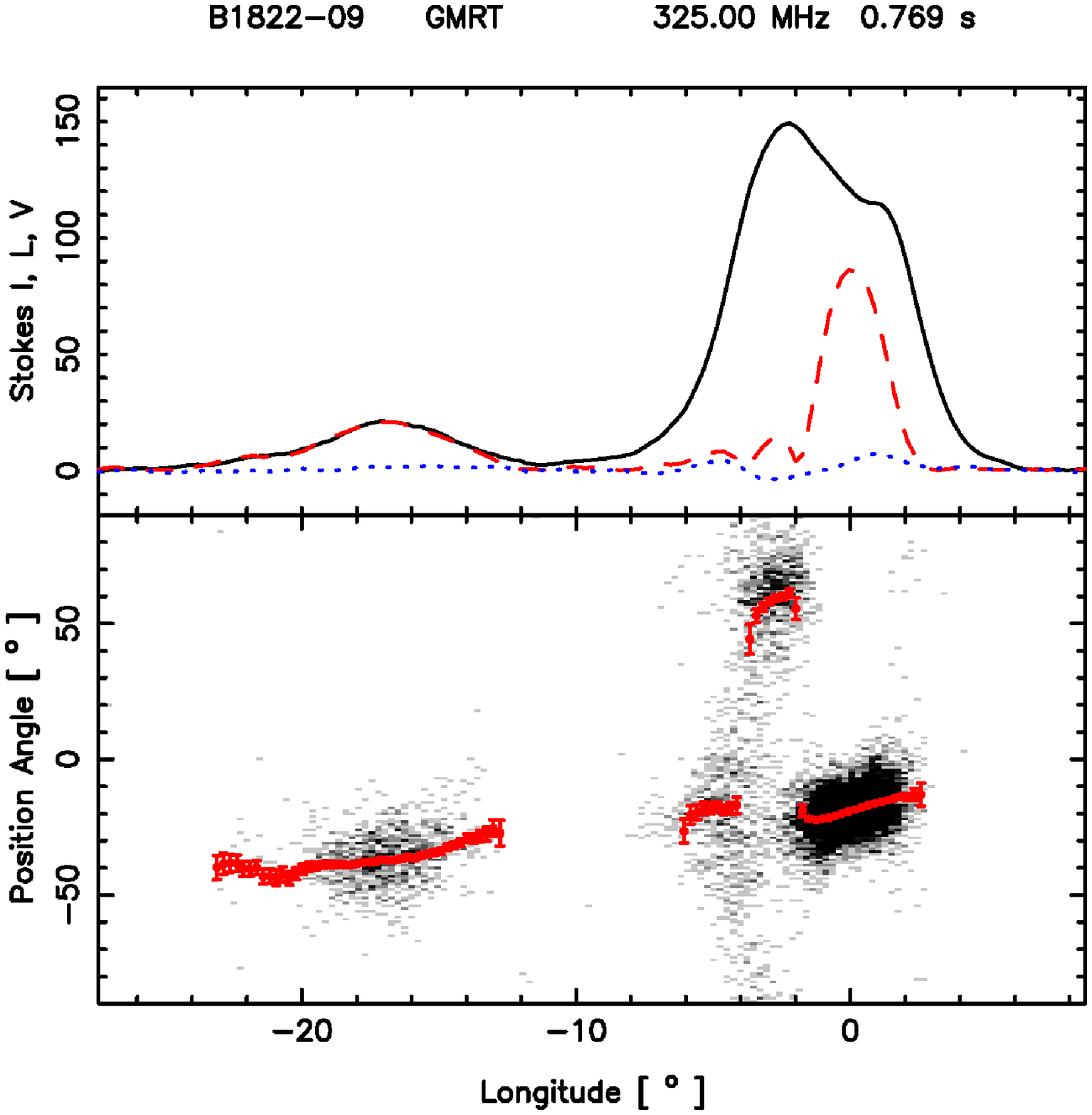}}}
{\mbox{\includegraphics[width=70mm,angle=-90,origin=c]{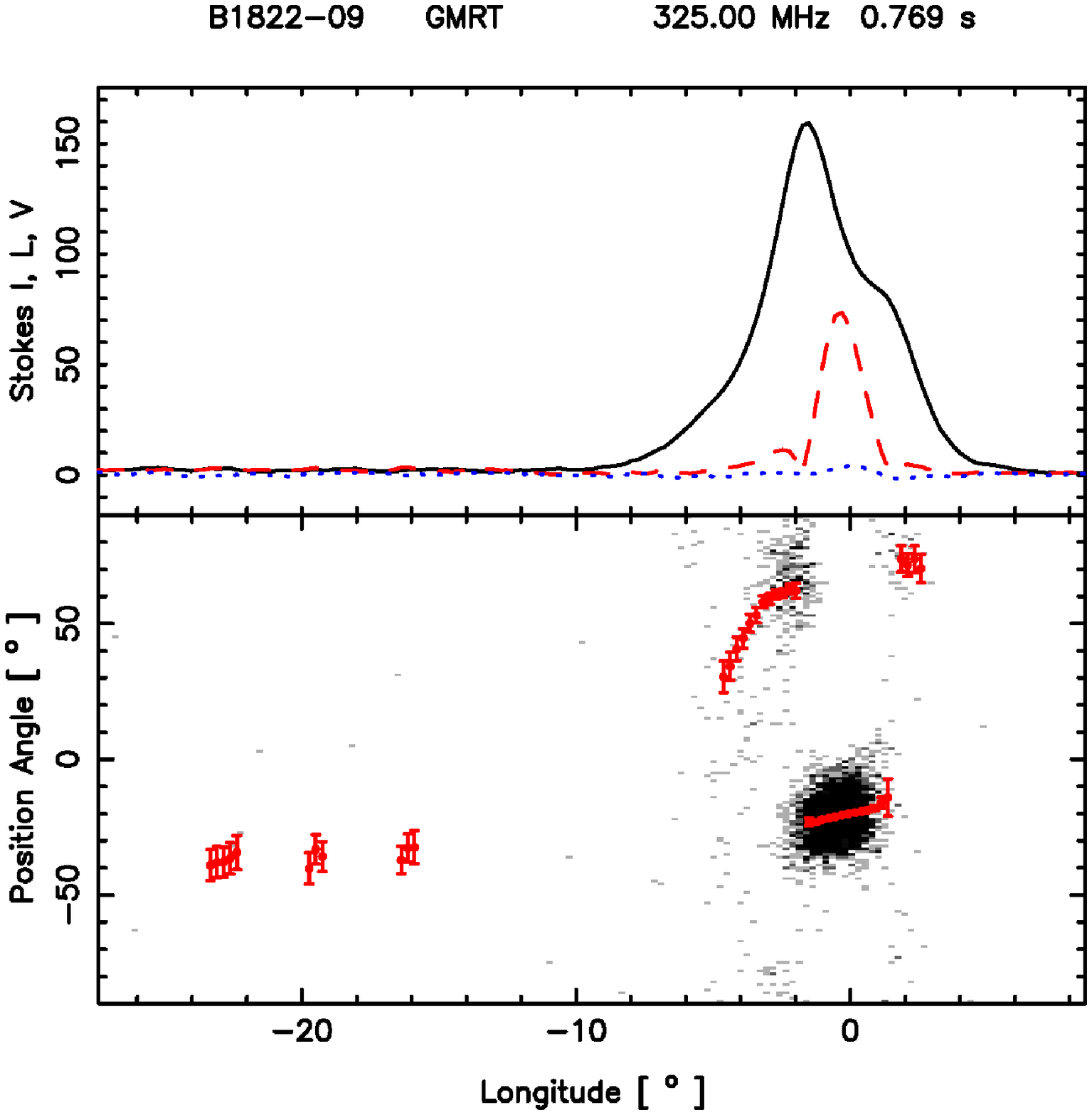}}}
%\end{tabular}
\caption{\psr\ 2077-pulse sequence (top left panel) observed with the GMRT 
at 325 MHz and used for analysis by \citet{backus2010}; note that a 
140$^{\circ}$ longitude interval is removed at the dashed vertical line.  PPA histogram 
(top right panel) corresponding to this observation.  Zoomed in panels showing 
the B-mode (bottom left panel) and Q-mode PPA histograms, respectively.  
Note that the PPA histograms below the main pulse lying between about --10$^{\circ}$ 
and +8$^{\circ}$ longitude are similar for both the B and Q modes.}
\label{figA1}
\end{center}
\end{figure*}

%****************************End Figure******************************************************

\subsection{Fitting the RVM}

We are in possession of the high quality polarized single pulse P-band GMRT observation 
that was used by BMR10.  Here we use this observation to explore whether the PPA traverse 
can successfully be fitted by the rotating-vector model (RVM).  The top-left panel of 
Figure ~\ref{figA1} shows the single-pulse sequence of 2077 pulses while the top-right panel 
shows the PPA histogram (as a gray scale in the bottom panel) for these pulses.  
Only those PPAs are shown in the histogram where the single pulse linear polarization 
power exceeds five times the rms noise level. The red points with error bars show the 
average PPA traverse.  It is clear from this plot that the single pulse linear polarized power 
is strong for the MP and the PC, but weak for the IP.  The bottom left and right plots show 
the PC and MP regions of the PPA histogram for the B and Q mode respectively.  
Qualitatively, the PPA behavior for these two modes is seen to be similar; it is the 
PPA distributions within the two modes that changes. 

Fitting the RVM to these PPA traverses is definitely a challenge.  The original 
version of the RVM introduced by \citet{rc69}  was advanced 
by \citet[][hereafter BCW]{bcw91} where the first-order special relativistic 
effects of aberration and retardation (A/R) were included.  In the most general case, 
the RVM depends on 5 parameters namely $\alpha$ the angle between the rotation 
and dipolar magnetic axes, $\beta$ the impact angle between the magnetic axis and 
the observer's line of sight, the distance $r_{em}$ from the center of the neutron star, 
the fiducial phase $\phi_{\circ}$, and the arbitrary shift of the PPA at the fiducial phase $\psi_{\circ}$. 
\begin{eqnarray}
\begin{split}
\psi = \psi_{\circ} + \\
\rm{tan^{-1}} \left( \frac{sin(\alpha)sin(\phi-\phi_{\circ})- (6\pi/P)(\rm{r_{em}/c)}sin(\eta)}
{sin(\eta)cos(\alpha) - sin(\alpha)cos(\eta)cos(\phi - \phi_{\circ})}\right)
\end{split}
\label{eq1}
\end{eqnarray}

Here $\eta = \alpha + \beta$, $\psi$ is the PPA, $\phi$ is the pulsar phase, $P$ the pulsar period and $c$ the velocity 
of light.  The $r_{em}$ term in the numerator is due to the A/R effects discussed by BCW, 
and for $r_{em}=0$ the equation reduces to the usual RVM.
For normal pulsars like \psr\ it is well known that the radio emission originates 
typically at an altitude of about 1\%-5\% of the light-cylinder radius, and the effect of A/R is 
to simply shift the PPA both in $\psi$ and $ \phi$, keeping the functional form of the 
traverse similar.  For the RVM fitting exercise below we will hence keep $r_{em} \sim 500$ km such 
that the quantity $(6\pi/P)(\rm{r_{em}}/c) \sim 0.04$, where $P=0.769$ sec. The usual strategy 
for fitting Eqn.~\ref{eq1} is to guess values for $\phi_{\circ}$ and $\psi_{\circ}$, based on the 
PPA traverse, and search for the parameters $\alpha$ and $\beta$.  Unfortunately, for a given 
set of $\alpha$ and $\beta$ the RVM can only be distinguished at the wings of the main pulse 
provided the width of the MP exceeds typically beyond 30$^{\circ}$. Certainly this is not the 
case for \psr, where the MP is only about 12$^{\circ}$ wide.  This essentially means 
that fitting the RVM will not yield meaningful constrains on $\alpha$ and $\beta$.

Let us illustrate the above point by doing the fits.  We first choose the Q-mode profile 
where the effect of the PC is low and the MP is prominent.  At this stage we should 
ask why the PC is not included in the RVM fits, and the reason is that the PC observed 
in several pulsars appears to be a different emission component with no clear polarizational 
relationship to that of the MP.  The shape of pulse profiles can usually be explained by 
the core-cone model where a central core emission component is surrounded by pairs of 
nested conal features.  Cones are of two types, outer and inner.  Outer cones show 
radius-to-frequency mapping (RFM), whereas the inner cones evolve with frequency 
much less than do outer cones.  In a detailed study of pulsars with PC components by 
\citet{bmr15}, the authors demonstrated that the location of the PC 
components w.r.t. the MP remain constant with frequency, whereas the components 
in the MP usually show the expected radius to frequency mapping.  For \psr\ the 
PC-MP separation (estimated as the peak of the PC and that of the MP) is about 
14.3$^{\circ}$ over the band from 243 MHz to 10.3 GHz, whereas the width (estimated 
using half power
point) of the MP evolves from 8.1$^{\circ}$ to 3.6$^{\circ}$ over this frequency range 
(see their Fig 1. and Table 2). 
Constancy of width with frequency is also a known property of inner cones within 
the core-cone model \citep{mr02}.  (PSR B1822--09's leading PC component 
might be modeled as an outer cone where the usual trailing component is not seen; 
however, its lack of evolution with frequency strongly mitigates against this explanation.)  
The MP on the other hand can be classified as a triple under the core-cone model 
showing the usual inner cone lack of RFM.  We will also not include the IP for fitting the 
RVM.  There are several reasons for doing so, but the most important one is that the 
sensitivity of our data does not allow us to interpret the PPA traverse by disentangling its
Orthogonal Polarization Mode behaviour.

\begin{figure}
\begin{center}
{\mbox{\includegraphics[width=62mm,angle=-90]{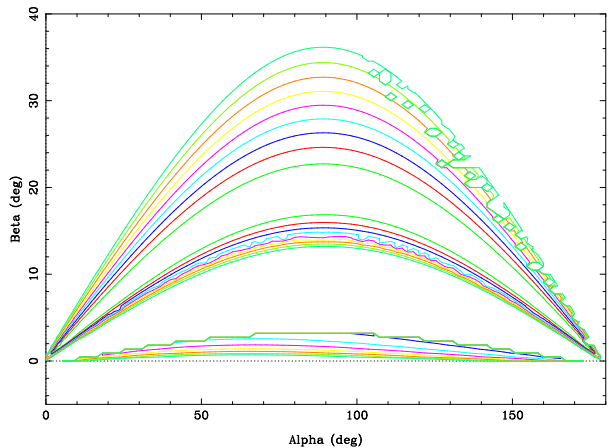}}}
\caption{The $\chi^{2}$ contours are shown of fitting the RVM for various values 
of $\alpha$ along the x-axis and $\beta$ along the y-axis.  Clearly the $\chi^{2}$ minimum is 
unconstrained. The curves represent multiples of the minimum $\chi^2$ 
	as 2 (green), 3 (red), 4 (blue), 5 (cyan), 6 (magenta) and 7 (yellow), 
	8 (orange), 9 (yellow-green), 10 (cyan-green).}
\label{fig2}
\end{center}
\end{figure}

\begin{figure*}
\begin{center}
{\mbox{\includegraphics[width=78mm,angle=-90]{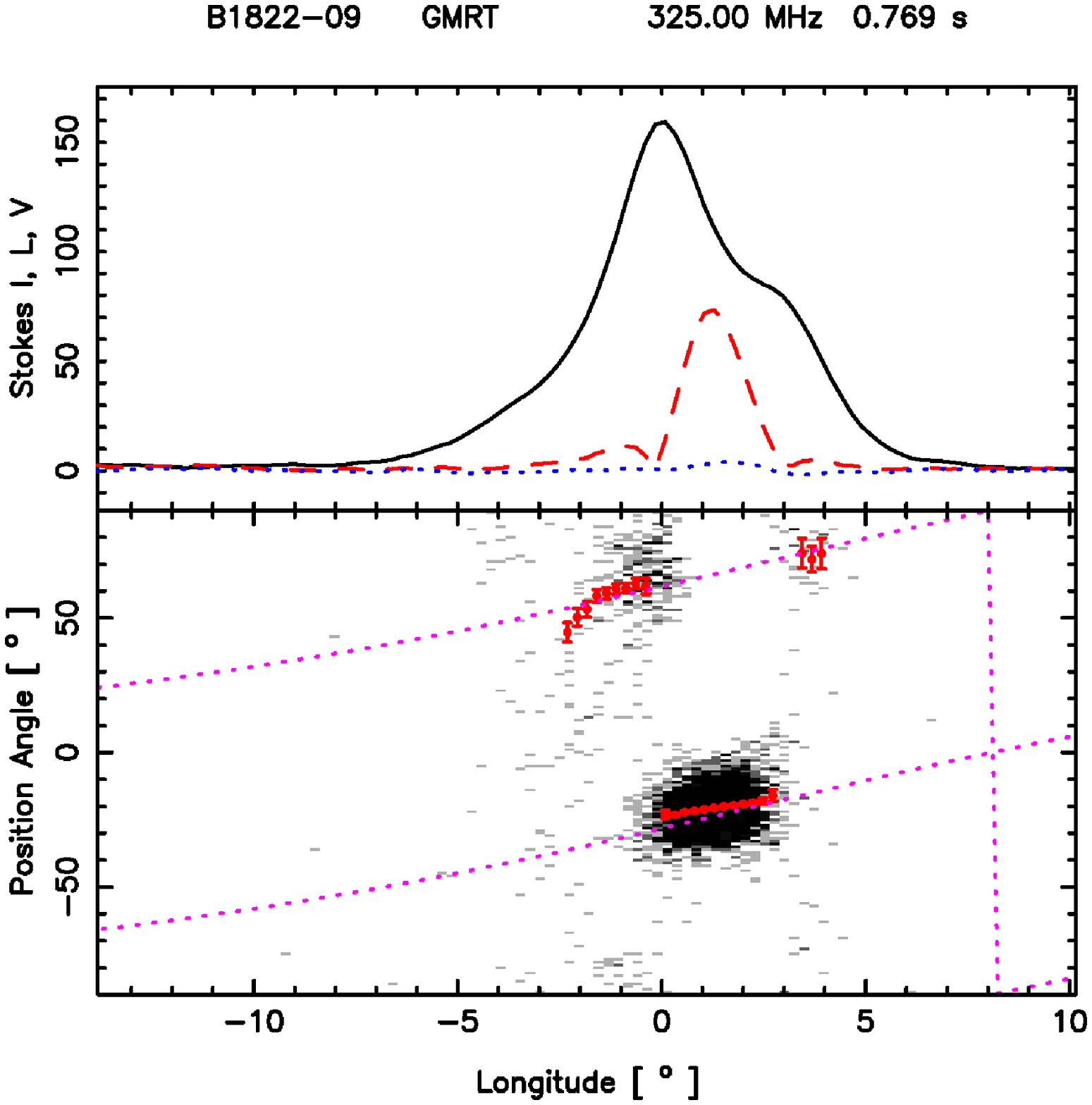}}} 
{\mbox{\includegraphics[width=78mm,angle=-90]{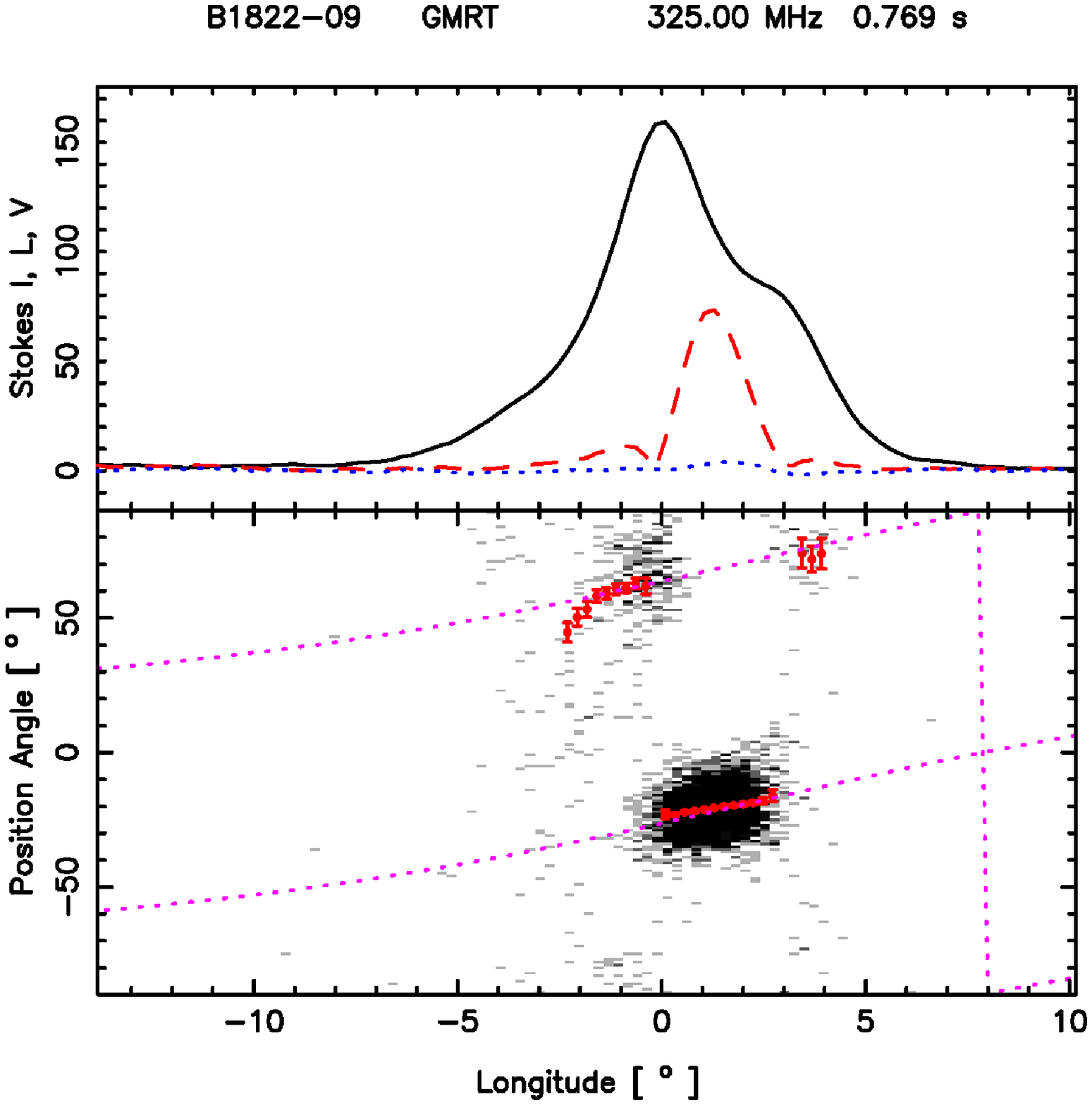}}}
\caption{The left and right panels show the RVM fits (as dashed magenta lines) 
where the left solution corresponds to an approximately orthogonal case 
with $\alpha = 84^{\circ}$ and $\beta = 16^{\circ}$ and the right solution corresponds 
to an almost aligned case with $\alpha = 7^{\circ}$ and $\beta = 1^{\circ}$. These 
two fits cannot be distinguished.}
\label{fig3}
\end{center}
\end{figure*}

Returning to fitting the RVM to the MP emission, we need to provide initial values 
for $\phi_{\circ}$ and $\psi_{\circ}$.  We assign $\phi_{\circ}$ (and read off the 
corresponding $\phi_{\circ}$ value) as the phase corresponding to the peak of the 
Q-mode profile, which is also the peak of the central core component of the triple 
profile classified by BMR10.  The PPA traverse at this phase shows an orthogonal 
jump, which is also characteristic of core emission.  With these values a grid search 
was carried out to find $\alpha$ and $\beta$, and the $\chi^{2}$ contours are shown in 
Fig.~\ref{fig2}.  Clearly a wide range of $\alpha$ and $\beta$ 
values are consistent with the fits, as the parameters are correlated up to 99\%.
The left and right panels of Fig.~\ref{fig3} show the RVM solutions for 
orthogonal and aligned rotator cases, respectively, and these solutions cannot 
be distinguished (see captions for more details).  Hence, RVM fits to the MP PPA 
traverse cannot constrain the geometry, in particular to distinguish between an 
orthogonal and aligned rotator.

\begin{figure*}
\begin{center}
\begin{tabular}{@{}lr@{}}
{\mbox{\includegraphics[width=78mm,angle=-90]{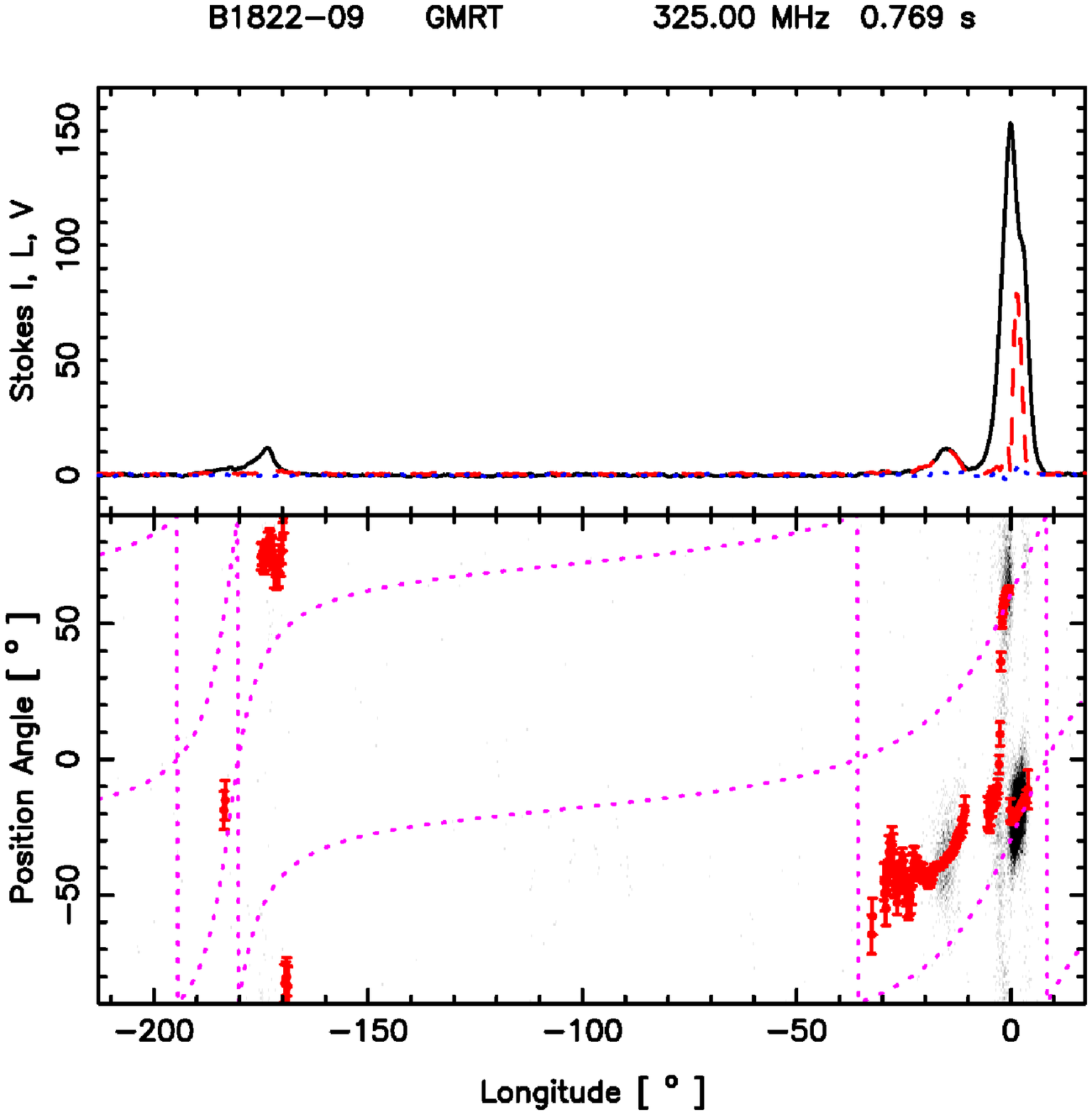}}}& \ \ \ \ \ \
{\mbox{\includegraphics[width=78mm,angle=-90]{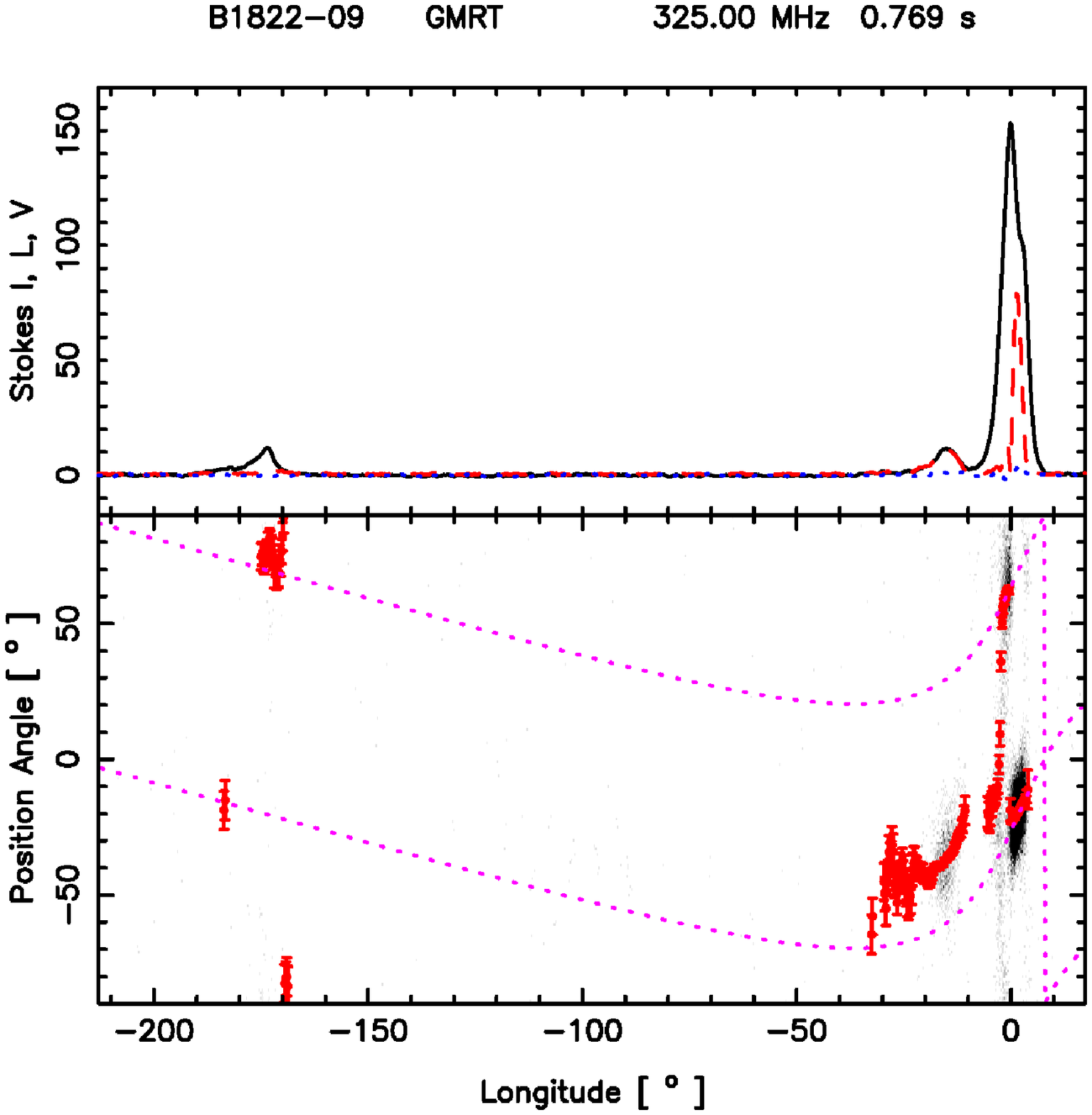}}}\\
\end{tabular}
\caption{The left panel in this figure shows the full IP-MP RVM for the almost orthogonal rotator case
using $\alpha = 84^{\circ}$ and $\beta = 16^{\circ}$. The right panel shows the full RVM for an almost 
aligned case 
using $\alpha = 7^{\circ}$ and $\beta = 1^{\circ}$. Note that in the IP region the two curves show 
very different behaviour. However since we cannot distinguish the OPM due to lack of sensitive 
data, we cannot trace the behaviour of the PPA traverse. The average PPA cannot be used for understanding
the RVM.}
\label{fig4}
\end{center}
\end{figure*}

If the IP could be included in the fits, there would be a possibility to distinguish 
between the aligned and orthogonal rotator geometries, since the PPA traverses 
are very different in the IP region as shown in Fig.~\ref{fig4}.  However, 
the lack of sufficiently sensitive single pulse polarization makes this task impossible for the moment.

\subsection{Geometry using core width}

If a core component can be identified in a pulse profile, then \citet{R90, R93a, R93b}
has shown that the half-width of the core component ($W_c$) can be related 
to $\alpha$ as $\sin(\alpha) = 2.45^{\circ}P^{-0.5}/W_c$.  Furthermore, using the 
steepest gradient point of the PPA traverse $R = \sin(\alpha)/\sin(\beta)$, $\beta$ 
can be found.  \citet{R93b} classifies the MP of PSR B1822--09 as having a 
core-cone configuration.  Furthermore, her Table 5 gives a value of 2.8$^{\circ}$ 
for $W_c$, implying per the above relationship that $\alpha \sim 86^{\circ}$.  Also 
a value of $R= 50$ \citep{LM88} has been used to get 
$\beta \sim 1.1^{\circ}$.  From our analysis we confirm that $W_c \sim 2.8^{\circ}$, 
however based on our best RVM solutions we find a significantly smaller value 
for the steepest gradient---i.e. $R \sim 2.4$ giving 
$\beta \sim 16^{\circ}$. The unusually large $\beta$ value is however worrying, 
since it leads to an unusually large beam radius of about 16$^{\circ}$ and emission 
heights of about 1300 km. However one should realize that for this pulsar the 
PPA histograms do not constrain $R$ very well.  It hence appears that the lack 
of a clean description of the RVM for \psr\ leads to an uncertain geometrical interpretation.

\subsection{Orthogonal Rotator}

While the methods for determining $\alpha$ and $\beta$ largely fail for \psr, 
there is striking evidence that the IP and the MP are separated by almost precisely 
180$^{\circ}$ at 325 MHz \citep[BMR10,][]{latham2012}
and nearly that over a broad 
band \citep{hankins1986} -- though it is important to restudy this separation
on the basis of modern techniques.  Furthermore, the narrowness of the MP core 
component is also strong evidence for an orthogonal alignment, the difficulty of 
determining $\beta$ notwithstanding.  Most recently, \citet{pilia2016} show 
in their Figure B.2 that 350 MHz and 1400 MHz profiles
are very well aligned, appearing to be the strongest argument for an orthogonal
rotator. Therefore, the existing evidence strongly 
points to \psr\ as having an orthogonal rotator geometry.  Of course, this 
orthogonal geometry raises an urgent question about the origin of the 
modulational ``cross-talk'' observed between the pole and interpole of 
\psr\ and many more MP-IP pulsars \citep[e.g.][]{WWS07, wwj12}.

\label{lastpage}

\end{document}